\documentclass[
twocolumn,
aps,
prb,
10pt,
longbibliography
]{revtex4-2}
 
\usepackage{amsmath,amssymb}
\usepackage{graphicx}
\usepackage{booktabs}
\usepackage[percent]{overpic}

\usepackage{hyperref}
\usepackage{cleveref}
\usepackage{dcolumn}
\usepackage{bm}
\usepackage{xr}
\usepackage{xr-hyper}

\begin{document}

\title{Strain-Tunable High-$T_c$ Superconductivity in Layered R3m B$_2$CN}

\author{Durjay Dhali}

\author{Arafat Rahman}

\author{Tareq Mahmud}
\email{tareqmahmud@du.ac.bd}

\author{Alamgir Kabir}
\email{alamgir.kabir@du.ac.bd}

\affiliation{Department of Physics, University of Dhaka, Dhaka 1000, Bangladesh}

\begin{abstract}
Using DFPT and Wannier-interpolated matrix elements in EPW, we find that isotropic tensile strain of 0--6\% narrows the B:2$p$/C:2$p$ bands at the Fermi level while softening the phonon branches most strongly coupled to them, driving the coupling constant $\lambda$ from 0.967 to 2.620. All four strained configurations remain mechanically and dynamically stable, satisfying the Born-Huang criteria with no imaginary phonon modes at any strain level. The fully anisotropic Migdal-Eliashberg equations give a $T_c$ that rises monotonically from 53.5~K to 84.5~K across the strain series, crossing the liquid-nitrogen boiling point at 6\% strain. The SISSO-based machine-learning model in EPW agrees with the anisotropic Migdal–Eliashberg $T_c$ values to within 10~K across the strain series.

\end{abstract}

\maketitle

\section{Introduction}
\label{sec:intro}

Conventional phonon-mediated superconductors have a key advantage over unconventional ones. Their properties can be described quantitatively from first principles~\cite{giustino2017electron,flores2020perspective}. Cuprates and iron-based compounds reach high transition temperatures, but their pairing mechanisms remain questioned~\cite{keimer2015quantum,paglione2010high}. Bardeen-Cooper-Schrieffer (BCS) theory~\cite{bardeen1957theory,bardeen1957microscopic} and its strong-coupling Migdal-Eliashberg extension~\cite{migdal1958interaction, eliashberg1960interactions} describe phonon-mediated materials with no such ambiguity. Advances in density-functional perturbation theory (DFPT)~\cite{baroni2001phonons} and Wannier-based electron-phonon interpolation~\cite{giustino2007electron} have made this theoretical framework more predictive. Computation-driven discovery of new high critical temperatures ($T_c$) phonon-mediated superconductors is now a realistic goal~\cite{pickard2020superconducting}.
High $T_c$ in conventional superconductors tends to arise from a combination of covalent bonding and light constituent elements~\cite{cerqueira2024sampling, kortus2001superconductivity}. Light atoms push phonon frequencies up, while covalent states near the Fermi level support strong electron-phonon matrix elements, a principle articulated early by Ashcroft~\cite{ashcroft1968metallic, ashcroft2004hydrogen}.  MgB$_2$ is the closest ambient-pressure case, with $T_c = 39$~K driven by coupling between boron $\sigma$-bands and bond-stretching phonon modes~\cite{nagamatsu2001superconductivity}. More recently, hydrogen-rich clathrates under pressure extend the logic further: H$_3$S reaches $T_c \approx 203$~K~\cite{drozdov2015conventional} and LaH$_{10}$ reaches $T_c \approx 250$~K~\cite{drozdov2019superconductivity}, alongside related hydride phases reaching comparable or higher $T_c$~\cite{duan2014pressure}. CaH$_6$ independently reaches $T_c \approx $ 210-215~K~\cite{li2022superconductivity,ma2022high}. However megabar pressures are required, and that constraint has pushed attention back toward ambient-pressure architectures built from light elements. Recent high-throughput searches have begun mapping how far conventional superconductivity can reach at ambient pressure~\cite{sanna2024prediction,gao2025maximum}.

Ternary boron-carbon-nitrogen compounds have both ingredients that drive high $T_c$ in conventional superconductors: light elements and strong covalent bonding~\cite{liu1989atomic}. B--C--N materials are exceptionally stiff, carry large Debye temperatures, and cover electronic behavior from insulating cubic BN to semimetallic graphite networks~\cite{tateyama1997proposed, solozhenko2001synthesis,knittle1989experimental,watanabe2004direct,bekaert2017evolution}. Diamond-like P3m1 B$_2$CX (X=N,P) phases have been predicted to reach $T_c$ near 46~K through coupling of metallic $\sigma$ electrons to B-C stretching modes~\cite{wan2024prediction}. BC$_5$, the diamond-like boron-carbide analog, was predicted computationally~\cite{calandra2008high} and subsequently synthesized~\cite{solozhenko2009ultimate}. Hard BN clathrates have been predicted to superconduct~\cite{li2019hard} and H or Li-doped cubic BN was recently predicted to reach $T_c$ above 120~K at ambient pressure~\cite{ding2024hydrogen}. More recently, anisotropic Migdal-Eliashberg calculations on the related superhard ternaries B$_2$C$_3$N and B$_4$C$_5$N$_3$ found $T_c \approx$ 20--40~K using the same EPW-based methodology adopted here~\cite{smith2026first}. Layered metal borocarbides have separately been proposed to reach $T_c$ up to 73~K via anisotropic Migdal-Eliashberg analysis~\cite{tomassetti2024prospect}. Despite the growing interest in the B--C--N family, R3m B$_2$CN has received comparatively little attention as a superconducting candidate.

R3m B$_2$CN is metallic, with Fermi-level states dominated by B and C p-orbitals a layered quasi-2D covalent framework~\cite{saha2020high}. That combination of light atoms, stiff bonds, and metallic p-states is what the Ashcroft-MgB$_2$ model~\cite{ashcroft2004hydrogen,nagamatsu2001superconductivity} points toward for strong electron-phonon coupling. Stiff covalent bonding raises phonon frequencies but lowers $N(E_F)$ through highly dispersive bands~\cite{ashcroft1976solid}. Lattice expansion offers a route to enhance electron-phonon coupling without changing chemical composition~\cite{locquet1998doubling,PhysRevLett.93.147006,henriquez2024modulation,johansson2022effect}. In light-element covalent compounds, the mechanism is especially direct, since phonon softening and band narrowing reinforce each other. R3m B$_2$CN satisfies both conditions, yet the strain dependence of its superconductivity has not been characterized. Reaching $T_c$ above 77~K at ambient pressure is also practically relevant, since liquid-nitrogen cooling is far cheaper and simpler than the liquid-helium or megabar-pressure requirements of other high $T_c$ routes~\cite{drozdov2015conventional,drozdov2019superconductivity}.
In this work, we report a first-principles study of R3m B$_2$CN under isotropic tensile strains of 0\%, 2\%, 4\%, and 6\%. We examine structural stability, electronic structure, lattice dynamics, and superconducting properties. The motivation of this work is to use isotropic tensile strain as a viable tuning parameter to enhance the superconducting properties of R3m B$_2$CN.

\section{Computational methods}
\label{sec:method}
Our density functional theory (DFT)~\cite{kohn1965self,hohenberg1964inhomogeneous} calculations were performed using the Quantum ESPRESSO (QE)~\cite{giannozzi2009quantum,giannozzi2017advanced,giannozzi2020quantum} package. Electron-ion interactions were described using ultrasoft pseudopotentials (USPP)~\cite{vanderbilt1990soft}. Exchange-correlation effects were treated with the Perdew-Burke-Ernzerhof (PBE) functional under the generalized gradient approximation (GGA)~\cite{perdew1996generalized,prandini2018precision}. Kohn-Sham wavefunctions were expanded in a plane-wave basis at a kinetic energy cutoff of 60 Ry, with the charge density cutoff set to 480 Ry. A Gaussian broadening scheme~\cite{elsasser1994density} with a smearing width of $\sigma$ = 0.02 Ry was applied to handle the metallic occupancies of B$_2$CN. NSCF calculations used a $\Gamma$-centered $24\times24\times2$ Monkhorst-Pack mesh~\cite{monkhorst1976special}.  Total and atom-projected densities of states used the tetrahedron~\cite{blochl1994improved} method on this k-mesh. Lattice dynamics were computed using DFPT as implemented in ph.x. Dynamical matrices were calculated on a $6\times6\times2$ q-mesh. These meshes reflect the quasi-two-dimensional geometry of the layered structure. Band structures were computed along    $\Gamma$--T--H$_2|$H$_0$--L--$\Gamma$--S$_0|$--S$_2$--F--$\Gamma$    of the hexagonal Brillouin zone from a non-self-consistent calculation on the converged charge density. Fermi surfaces were visualized using FermiSurfer~\cite{kawamura2019fermisurfer}. 

To examine the effect of strain on the superconducting properties, isotropic tensile strain was applied to the equilibrium R3m structure by uniformly scaling the lattice parameters while preserving the crystal symmetry. Fixing the internal coordinates isolates the volumetric contribution to the electron-phonon response from any competing ionic relaxation. The mechanical and dynamical stability of each strained configuration was confirmed independently via Murnaghan equation of state~\cite{murnaghan1944compressibility,birch1947finite} and phonon calculations. The equation of state (EOS) and elastic tensor for each strained configuration were computed using the thermo\_pw package~\cite{wu2005systematic,thermo_pw}. For the EOS, total energies at a set of uniformly scaled volumes were fitted to the Murnaghan equation of state to extract the equilibrium volume, bulk modulus, and its pressure derivative.

The elastic tensors were computed for each strained configuration using the stress-strain approach as implemented in thermo\_pw. Mechanical stability was checked against the Born-Huang criteria ~\cite{born1996dynamical}. Various elastic parameters were calculated using ELATE: an elastic tensor analysis tool~\cite{gaillac2016elate}. The Debye temperature $\Theta_D$ was calculated from the VRH-averaged~\cite{hill1952elastic} sound velocities via Anderson's procedure ~\cite{anderson1963simplified}.  

Electron-phonon matrix elements and superconducting properties were computed with the EPW code~\cite{noffsinger2010epw,ponce2016epw,lee2023electron}. EPW combines maximally localized Wannier functions (MLWFs)~\cite{marzari1997maximally,souza2001maximally} with Wannier interpolation to access dense Brillouin zone grids at modest computational cost. MLWFs were constructed via Wannier90~\cite{pizzi2020wannier90} using initial projections onto B-2p, C-2p, and N-2p orbitals. These projections capture the dominant orbital character at the Fermi level ($E_F$) in R3m B$_2$CN and give well-localized functions with small spread. NSCF calculations for the Wannierization step used a $\Gamma$-centered $24\times24\times4$ $k$-mesh. Coupling matrix elements were calculated on a coarse $12\times12\times2$ k-mesh and $6\times6\times2$ q-mesh, then interpolated onto fine $24\times24\times8$ k and $12\times12\times4$ q grids. The Coulomb pseudopotential~\cite{morel1962calculation} was set to $\mu^*$ = 0.10, consistent with values used for covalent light-element systems such as MgB$_2$~\cite{choi2002origin,kong2001electron}.  The Eliashberg spectral function~\cite{allen1972neutron} $\alpha^{2}F(\omega)$ and cumulative coupling~\cite{dynes1972mcmillan} $\lambda$ were also extracted at this fine grid. The Allen-Dynes temperature ($T_c^{\rm AD}$) was also evaluated from the interpolated $\alpha^2F(\omega)$ and logarithmic phonon frequency~\cite{allen1975transition} $\omega_{\rm log}$ to provide an isotropic reference alongside the anisotropic result, following the formula of Allen and Dynes.

The anisotropic Migdal-Eliashberg equations were solved on the Matsubara frequency axis as implemented in EPW~\cite{margine2013anisotropic}. Unlike the Allen-Dynes formula, this approach retains the full k-point dependence of the gap function $\Delta_{nk}(i\omega_m)$ and mass renormalization $Z_{nk}(i\omega_m)$ and has been shown to give transition temperatures closer to experiment in systems with pronounced Fermi surface anisotropy. R3m B$_2$CN falls in this category, with multiple bands crossing the Fermi level. The superconducting transition temperature ($T_c$) was estimated from the temperature dependence of the anisotropic gap, taking the gap-closing point as $T_c$. Gap distributions at each temperature were obtained from the k-resolved values $\Delta_{nk}$ on the fine mesh. The quasiparticle DOS and band-resolved coupling $\lambda_{nk}$ were computed from the same solution to characterize the pairing state across the strain series. As an independent check, $T_c$ was also estimated from the SISSO-based machine-learning model implemented in EPW, and the resulting $T_c^{\mathrm{ML}}$ values were compared against the Migdal-Eliashberg predictions across the strain series.

\section{Crystal structure \& stability}
\label{sec:stability}
\subsection{Crystal Structure and Strain Series}

\begin{figure}[]
  \centering
  \includegraphics[width=0.35\textwidth]{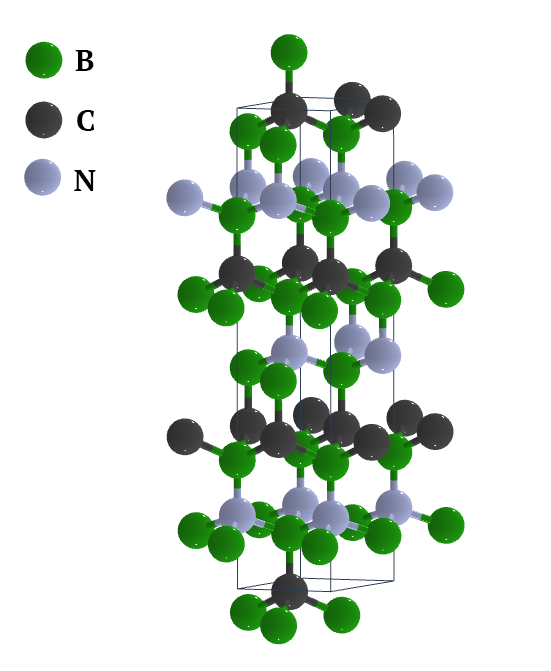}
  \caption{Crystal structure of B$_2$CN}
  \label{fig:structure}
\end{figure}

R3m B$_2$CN is a rhombohedral layered structure (space group No. 160) with 12 atoms per unit cell -- three formula units. B, C and N layers alternate along c-axis, connected by strong covalent B--C and B--N bonds, with each atom in a threefold-coordinate planar environment (Fig.~\ref{fig:structure}). The equilibrium lattice parameters are a=2.586~\AA~ and c=13.177~\AA, with c/a $\approx$ 5.09.

The Murnaghan EOS has the form,
\begin{equation}
	P(V)=\frac{B_0}{B_0'}\left[\left(\frac{V}{V_0}\right)^{-B_0'}-1\right]
\end{equation}
where $V_0$, $B_0$, and $B_0'$ are the equilibrium volume, bulk modulus, and its pressure derivative. The fit gives an equilibrium volume of V$_0$ = 76.40 \AA$^3$ at 0\% strain. The bulk modulus increases monotonically with strain - 320.0, 322.1, 330.5, and 347.5 GPa at 0\%, 2\%, 4\%, and 6\%. As bond lengths in the B-C-N network extend, the structure resists further volumetric compression more strongly. This is consistent with the behavior expected of a stiffening covalent framework.

\subsection{Mechanical Stability}

For the elastic constants, small symmetry-inequivalent strains were applied to the relaxed cell and the resulting stress tensors used to construct the full $C_{ij}$ tensor. The R3m group yields six independent constants $C_{11}$, $C_{12}$, $C_{13}$, $C_{33}$, $C_{44}$ and $C_{66}$ -- with $C_{66}=(C_{11}-C_{12})/2$ fixed by symmetry. The results for all four strain levels are collected in Table~\ref{tab:elastic_constants}.

\begin{table}[htbp]
\setlength{\tabcolsep}{8pt}
\centering
\begin{tabular}{lcccc}
\toprule
$C_{ij}$ (GPa) & $\varepsilon = 0\%$ & $\varepsilon = 2\%$ & $\varepsilon = 4\%$ & $\varepsilon = 6\%$ \\
\midrule
$C_{11}$      & 567.97 & 568.32 & 567.12 & 561.85 \\
$C_{12}$      & 225.39 & 225.61 & 224.86 & 221.51 \\
$C_{13}$      & 111.56 & 111.75 & 111.09 & 108.15 \\
$C_{33}$      & 851.44 & 851.85 & 850.44 & 844.19 \\
$C_{44}$      & 292.97 & 293.06 & 292.72 & 291.20 \\
$C_{66}$      & 171.29 & 171.35 & 171.13 & 170.17 \\
$B_0$         & 320.0  & 322.1  & 330.5  & 347.5  \\
$\Theta_D$    & 1555.0 & 1555.2 & 1554.4 & 1550.9 \\
$K_V$         & 320.49 & 320.74 & 319.86 & 315.95 \\
$E_V$         & 602.83 & 603.09 & 602.19 & 598.19 \\
$G_V$         & 254.04 & 254.12 & 252.83 & 253.52 \\
$\nu$         & 0.1865 & 0.1866 & 0.1862 & 0.1844 \\
$K_V/G_V$     & 1.26   & 1.26   & 1.26   & 1.26   \\
\hline
\end{tabular}
\caption{Elastic constants $C_{ij}$ (GPa),  Murnaghan bulk modulus B$_0$ (GPa), Debye temperature $\Theta_D$ (K), Voigt bulk modulus $K_V$ (GPa), Young's modulus $E_V$ (GPa), Shear modulus $G_V$ (GPa), Poisson ratio $\nu$, and Pugh ratio $K_V$/$G_V$ for R3m B$_2$CN at each strain level}
\label{tab:elastic_constants}
\end{table}

Mechanical stability was checked against the Born-Huang criteria for trigonal crystals~\cite{born1996dynamical,mouhat2014necessary}, which require: $C_{11} > \left|C_{12}\right|$, $C_{44}>0$, $\left( C_{11}+C_{12} \right)C_{33}> 2C_{13} ^2$ and $\left( C_{11}-C_{12} \right)C_{44}> 0$. All four conditions are satisfied at every strain level. The constants are largely insensitive to strain below 6\%: $C_{11}$, $C_{33}$ and $C_{44}$ each shifted by less than 0.2\% across 0-4\% range. At 6\% strain all components soften slightly -- $C_{11}$ by 1.1\% and $C_{33}$ by 0.9\% -- but the Born-Huang conditions remain satisfied. Brillouin light scattering (BLS) can measure bulk and surface acoustic modes in thin-film and polycrystalline specimens~\cite{carlotti1997comparative,carlotti2005elastic}. The computed $C_{ij}$ values reported here can therefore be compared with future BLS measurements to experimentally validate the predicted mechanical response of B$_2$CN under strain.

The elastic moduli place B$_2$CN firmly in the hard covalent ceramic class~\cite{clerc1998mechanical}. $K_V \approx$ 320~GPa and $E_V \approx$ 600~GPa  shift by less than 1.5\% across the full 0-6\% strain. This indicates that the B--C--N network is not meaningfully disrupted at these strain levels. At $\nu \approx$  0.186 the Poisson ratio is lower than typical transition-metal nitrides and carbides ($\nu \approx$  0.20-0.30)~\cite{kral1998critical} but sits above cubic BN ($\nu \approx$  0.12)~\cite{kastuar2023mechanical} and diamond ($\nu \approx$  0.148)~\cite{klein1993young}, consistent with the mixed ionic covalent character of the B--C--N framework. The Pugh ratio $K_V/G_V = $ 1.26 falls well below ductile-brittle threshold of 1.75~\cite{pugh1954xcii,chen2011modeling} at all strain levels, indicating that the material is intrinsically brittle across the series. The Debye temperature $\Theta_D$ is  $\approx$ 1555~K at 0\% strain and falls by just 0.3\% to 1551 K at 6\%. This places B$_2$CN well within empirical bound $T_C \leq A \Theta_D$ proposed for phonon-mediated superconductors~\cite{esterlis2018bound}, later validated across the Materials Project database using machine-learned $\Theta_D$ predictions~\cite{smith2023machine}. For comparison, cubic BN has $\Theta_D \approx$ 1700~K~\cite{grimsditch1994elastic} and MgB$_2 \approx$ 750~K~\cite{bud2001boron}. The high value in B$_2$CN follows from the light atomic masses and stiff B--C--N bonds.

\section{Electronic structure}
\label{sec:elec}

\begin{figure*}[]
  \centering
  \includegraphics[width=\textwidth]{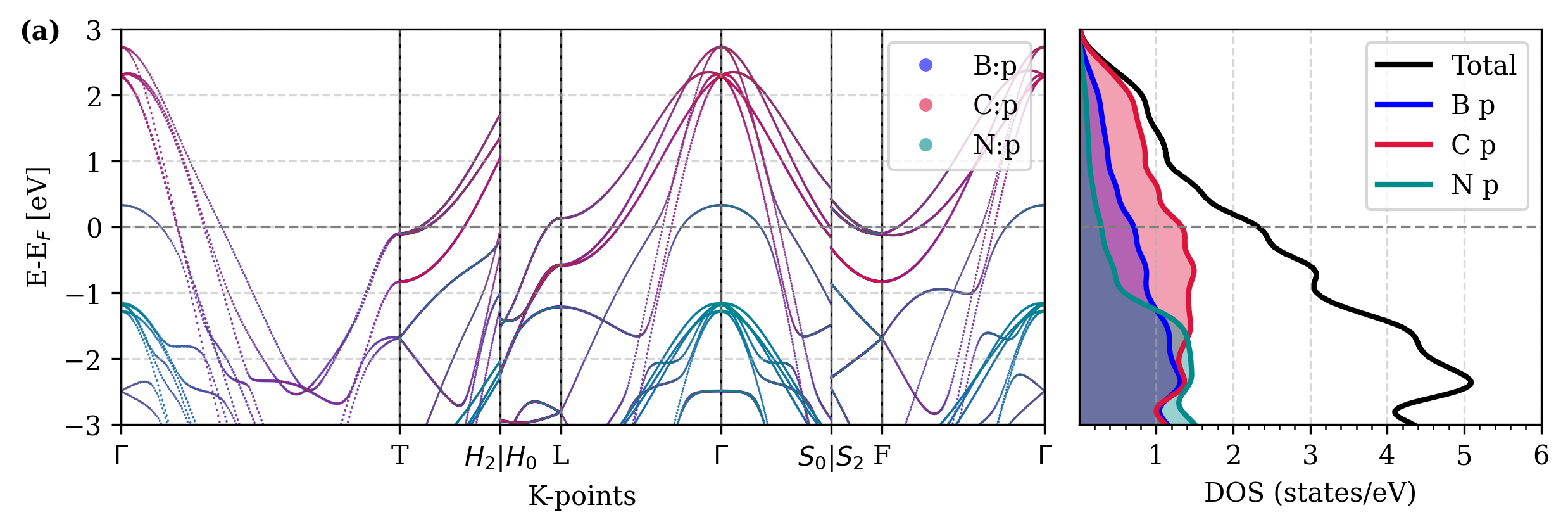}\\
  \includegraphics[width=\textwidth]{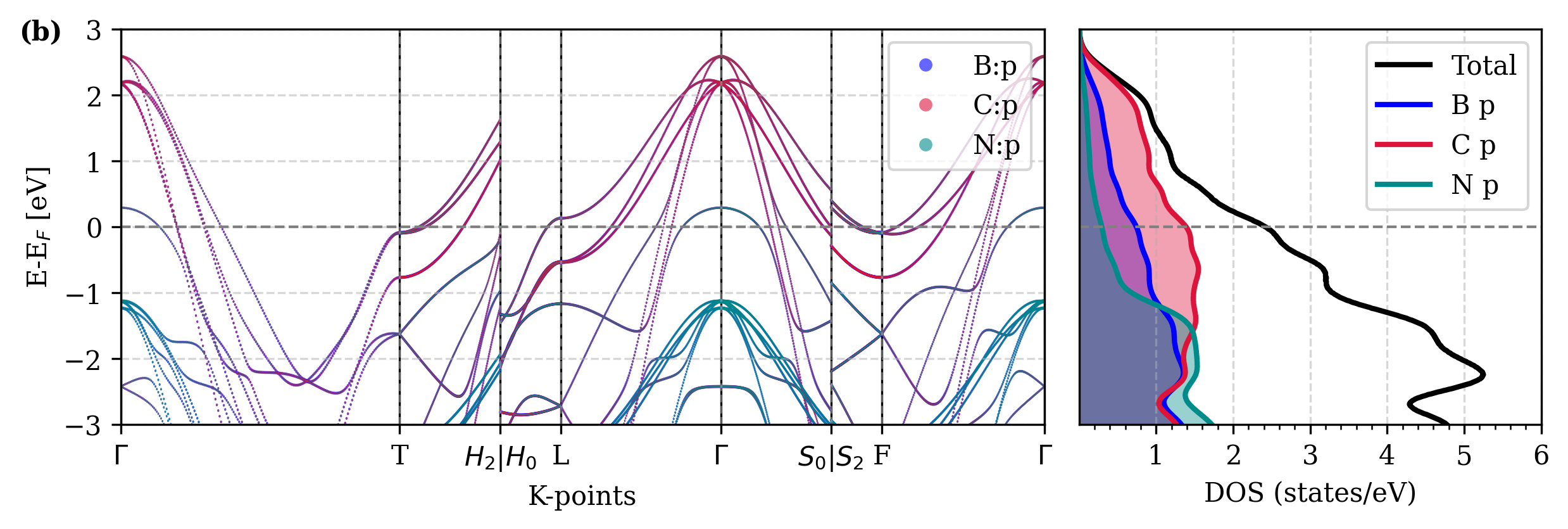}\\
  \includegraphics[width=\textwidth]{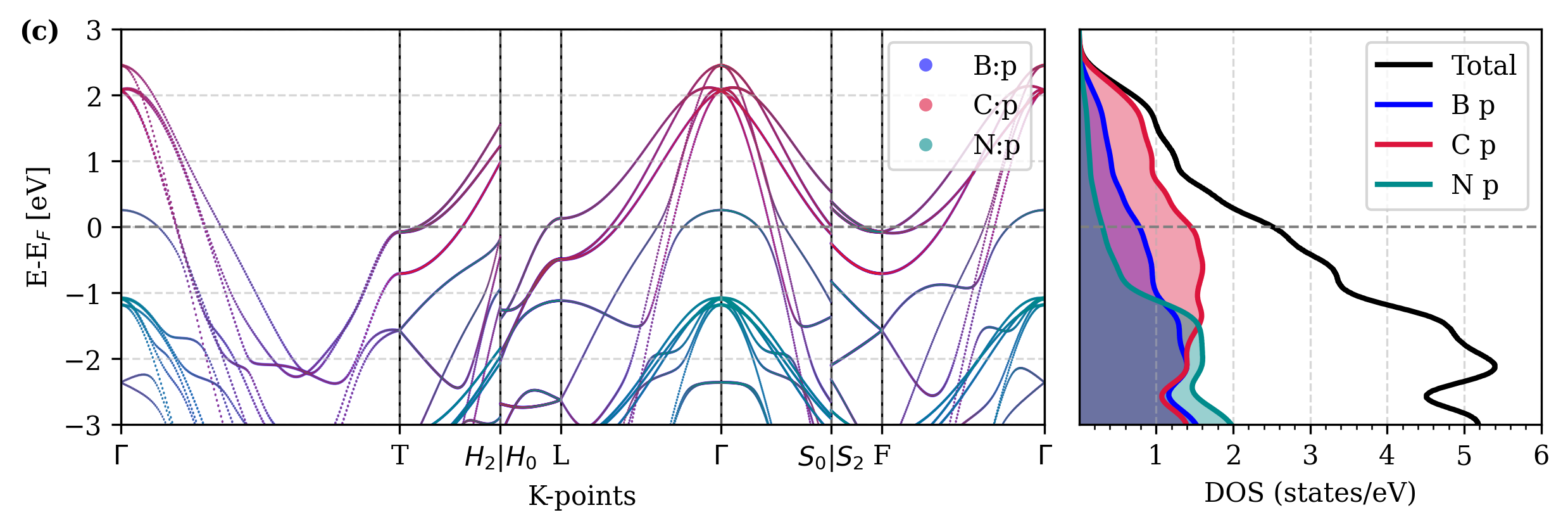}\\
  \includegraphics[width=\textwidth]{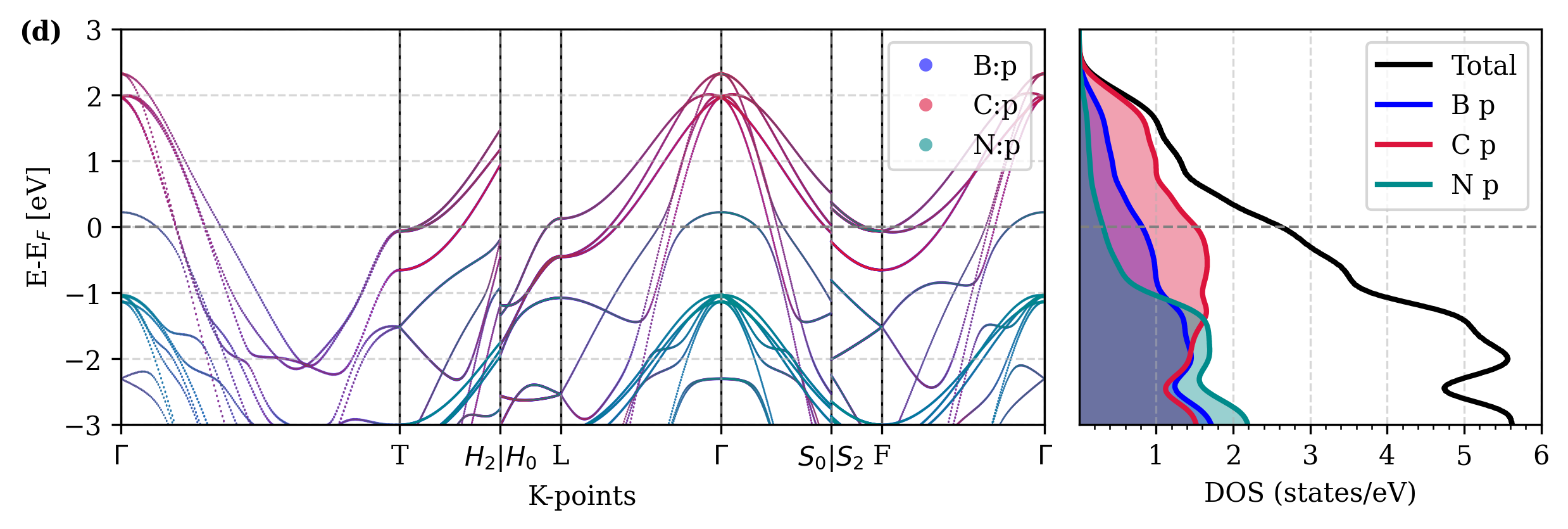}
  \caption{Electronic band structure and DOS for R3m B$_2$CN under isotropic tensile strain of (a) 0\%, (b) 2\%, (c) 4\%, and (d) 6\%.}
  \label{fig:bandstructure}
\end{figure*}

\begin{figure}[]
  \centering
  \includegraphics[width=0.5\textwidth]{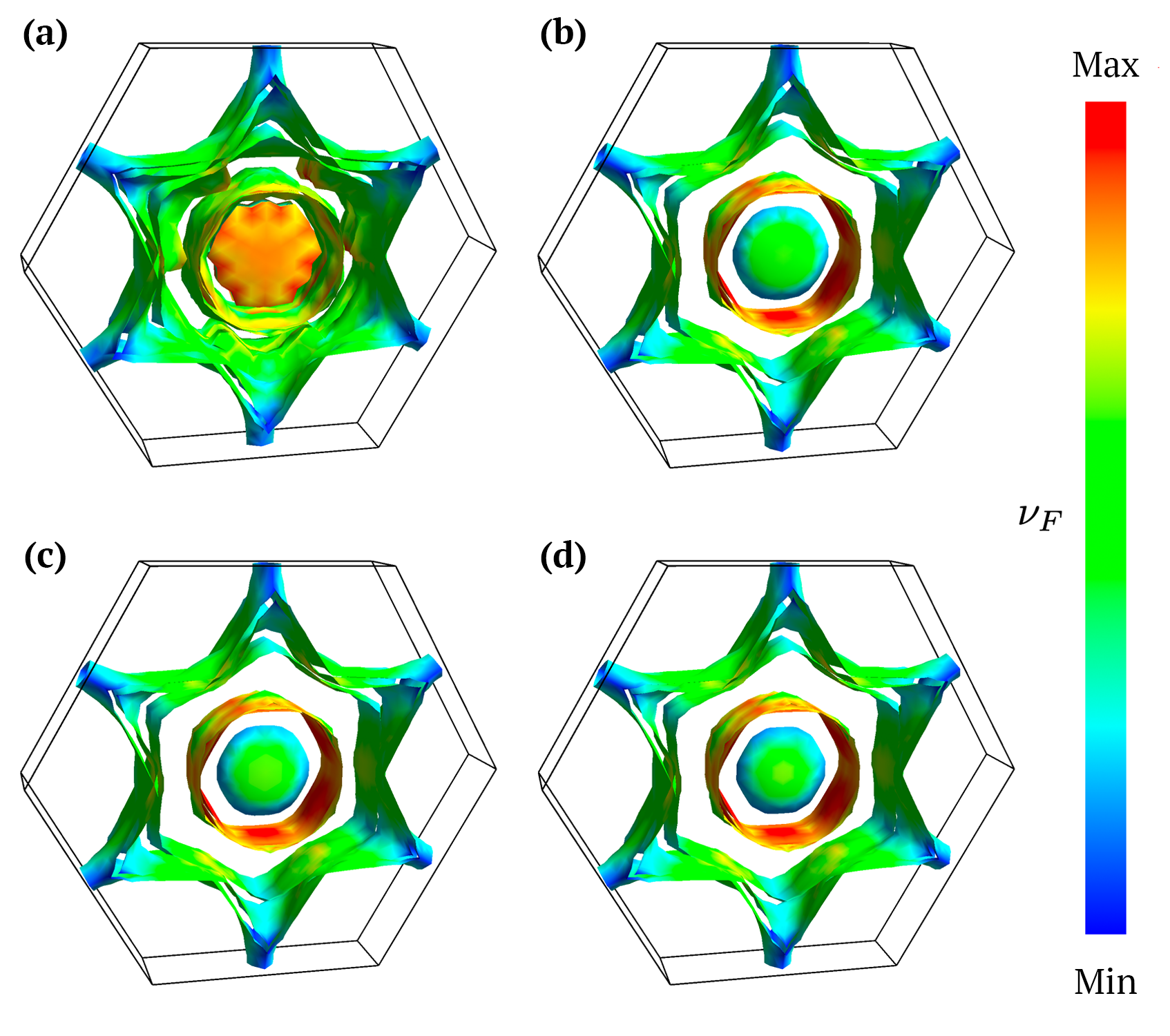}
  \caption{Fermi surface of R3m B$_2$CN under isotropic tensile strain of (a) 0\%, (b) 2\%, (c) 4\%, and (d) 6\% with the color mapping representing the magnitude of the Fermi velocity $\nu_F$.}
  \label{fig:fermisurface}
\end{figure}

The electronic band structure and atom-projected density of states (PDOS) R3m B$_2$CN at 0-6\% isotropic tensile strain are shown in Figure~\ref{fig:bandstructure}. The k-path follows the standard rhombohedral Brillouin zone:   $\Gamma$--T--H$_2|$H$_0$--L--$\Gamma$--S$_0|$--S$_2$--F--$\Gamma$. The compound remains metallic across the entire strain range, with multiple bands crossing the Fermi level at all four strain levels. Near $E_F$ the spectral weight is dominated by B:2$p$ and C:2$p$ states. N:2$p$ states lie largely below $E_F$, with a broad peak centered roughly 1-1.5 eV beneath it. These N:$p$ states form the occupied bonding network rather than contributing to metallic conduction. This follows from the relatively higher electronegativity of nitrogen compared to boron and carbon~\cite{pauling1932nature}. The Fermi surface is therefore defined by B and C character at all strain levels considered.

The orbital character of the Fermi-level crossing bands remains unchanged across the strain series. The same B:p and C:p bands cross $E_F$ at all strains from 0\% to 6\%. Fermi-level crossings fall in the same regions of the Brillouin zone, and the orbital character of the relevant bands is preserved. N:p bands remain predominantly occupied. No metal-to-insulator transition~\cite{mott1968metal}, Lifshitz transition~\cite{lifshitz1960anomalies,blanter1994theory}, or symmetry-lowering distortion is observed. R3m B$_2$CN is neither magnetically ordered nor symmetry-lowering. Isotropic tensile strain therefore acts as a clean perturbation, uniformly scaling the lattice without introducing additional complexity.

Bandwidth compression is the dominant electronic trend across the strain series~\cite{pereira2009strain}. As the lattice expands, orbital overlaps between neighboring B, C, and N sites weaken and effective hopping integrals decrease. The B:p and C:p bands near $E_F$ become progressively less dispersive. In Figure~\ref{fig:bandstructure},  the energy range spanned by the Fermi-level bands at 6\% is visibly narrower than at 0\%, while the Fermi crossings themselves remain intact. The bands narrow gradually, with no discontinuity. This smooth trend matches the way strain steadily reduces orbital overlap.

Band narrowing at $E_F$ concentrates states into a narrower energy window, so $N(E_F)$ rises monotonically with strain. The total DOS at $E_F$ increases from 2.33~states~eV$^{-1}$ at 0\% to 2.41, 2.52, and 2.63~states~eV$^{-1}$ at 2\%, 4\%, and 6\% strain, respectively. The DOS panels in Figure~\ref{fig:bandstructure} confirm this directly. The total DOS at $E_F$ increases at each successive strain level.  B:p and C:p contributions account for most of the increase, as both sit at the Fermi level. The N:p contribution shifts modestly with lattice strain but remains anchored below $E_F$, because these nitrogen states are strongly hybridized into deep bonding orbitals. As a result, it has little effect on $N(E_F)$.

Figure~\ref{fig:fermisurface} shows the three-dimensional Fermi surfaces of R3m B$_{2}$CN under isotropic tensile strains of (a) 0\%, (b) 2\%, (c) 4\%, and (d) 6\% within the first Brillouin zone of the rhombohedral lattice. The color scale represents the local magnitude of the Fermi velocity $\nu_{F}$, ranging from minimum values in blue to maximum values in red. At all strain levels, the Fermi surface contains two different types of sheets. The large central pocket around the $\Gamma$ point is electron-like and has a rounded and convex shape. It originates from bands that cross the Fermi level with upward dispersion near the zone center. This is also consistent with the band structure in Figure~\ref{fig:bandstructure}, where C-$p$ and B-$p$ dominated bands cross $E_{F}$ near $\Gamma$. At 0\% strain in panel (a), the central electron pocket is irregular and strongly anisotropic, with large regions of high Fermi velocity shown by the orange-red colors. This indicates strongly dispersive bands near the Fermi level. As the tensile strain increases to 2\% and 4\% in panels (b) and (c), the central electron pocket becomes smoother and more isotropic with a more rounded shape. The mainly green color indicates intermediate Fermi velocities and suggests that the anisotropy of the band dispersion around $\Gamma$ is reduced by strain. At 6\% strain in panel (d), the central pocket becomes more compact and nearly hexagonally symmetric with mostly low-to-intermediate $\nu_{F}$ values in the blue-green range. This suggests that the related bands near $\Gamma$ become less dispersive under larger tensile strain.

The outer star-shaped sheets extend toward the boundaries of the hexagonal Brillouin zone and form six arm-like features. These sheets are hole-like and originate from bands that cross the Fermi level with downward dispersion near the zone-boundary regions, including the T, L, and F points. This is consistent with the downward-dispersing bands observed near T and along the $\Gamma$-T and L-$\Gamma$ directions in Figure~\ref{fig:bandstructure}. At 0\% strain, the hole-like arms are broad and interconnected and show a wide range of Fermi velocities. The neck regions mainly show blue-to-green colors, while the outer parts change from green to yellow. This indicates that the hole-like surface contains states with different band curvatures and dispersions. With increasing tensile strain, the hole-like arms become smaller and narrower. The connecting regions between neighboring arms become thinner at 2\% and 4\% strain, while the overall volume of the hole-like sheets decreases. This suggests that the contribution of hole carriers at the Fermi level is reduced with strain. At 6\% strain, the hole-like arms are much smaller and appear less connected, with mainly low Fermi velocities shown by the blue-green colors. This suggests that tensile strain shifts the related hole bands away from the Fermi level and limits their contribution to a smaller region of momentum space.

Overall, the Fermi surface changes continuously with tensile strain. The electron-like pocket around $\Gamma$ becomes more regular and less anisotropic, while the hole-like sheets near the zone boundaries become smaller. At the same time, the variation in Fermi velocity across the surface is reduced, and the total density of states at the Fermi level is modified. These changes are directly relevant to the superconducting properties of B$_{2}$CN, since the density of states at $E_{F}$ and the Fermi velocity enter directly into the electron-phonon coupling and the superconducting transition temperature $T_{c}$ through the McMillan--Allen--Dynes framework.

The superconducting implications of this trend are discussed in Section~\ref{sec:tc}. According to the McMillan-Hopfield expression~\cite{mcmillan1968transition,hopfield1969angular}, $\lambda \propto \frac{N(E_F)\langle I^2 \rangle}{M\langle \omega^2 \rangle}$, where $\langle I^2 \rangle$ is the Fermi-surface-averaged electron-phonon matrix element, $M$ the atomic mass, and $\langle \omega^2 \rangle$ the mean-square phonon frequency. Because $\lambda$ scales directly with $N(E_F)$, the band narrowing established above is favorable for electron-phonon coupling. Tensile strain drives this rise in $N(E_F)$ purely by narrowing the B:p and C:p bands at the Fermi level. The metallic character and orbital composition of the Fermi surface, and the sheet count established in Fig.~\ref{fig:fermisurface}, remain unchanged. As shown in Section~\ref{sec:tc}, however, the rise in $N(E_F)$ is modest compared to the accompanying phonon softening, which is the dominant contributor to the strain-induced enhancement of $\lambda$ and $T_c$. Angle-resolved photoemission spectroscopy (ARPES) on strained samples can detect both the bandwidth narrowing shown in Fig.~\ref{fig:bandstructure} and the associated rise in $N(E_F)$~\cite{damascelli2003angle,yang2018visualizing}. ARPES's momentum resolution allows direct examination of the underlying electronic mechanism, providing an experimentally accessible way to test these predictions before superconductivity measurements.

\section{Phonon Properties}
\label{sec:phonon}

\begin{figure*}[]
  \centering
  \includegraphics[width=\textwidth]{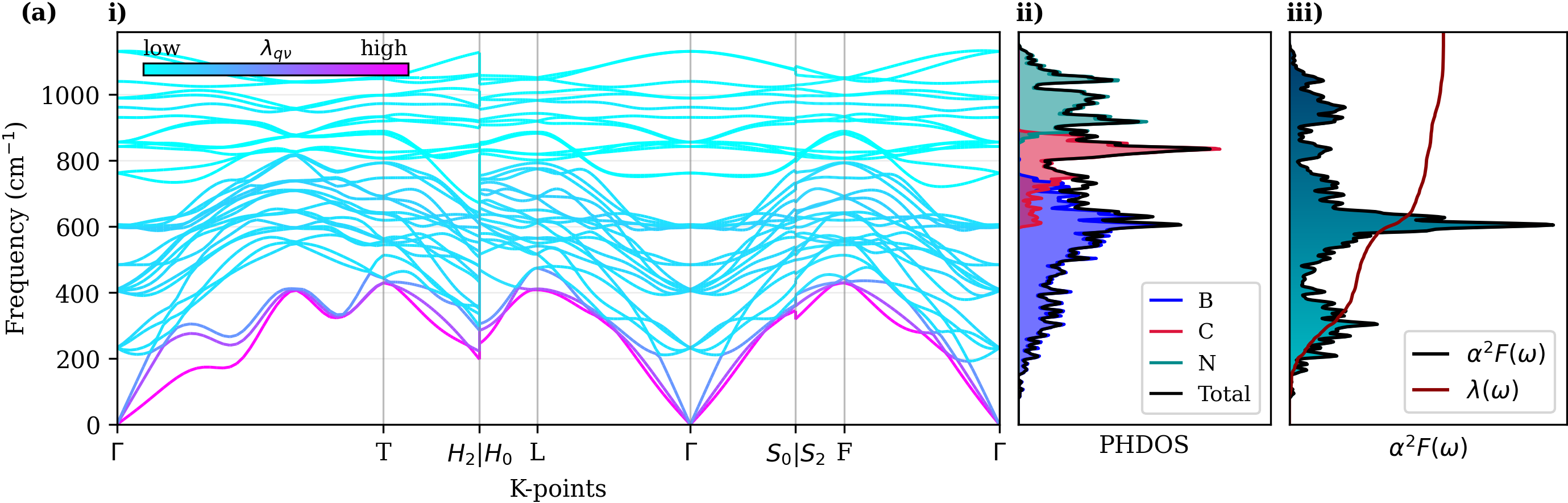}\\
  \includegraphics[width=\textwidth]{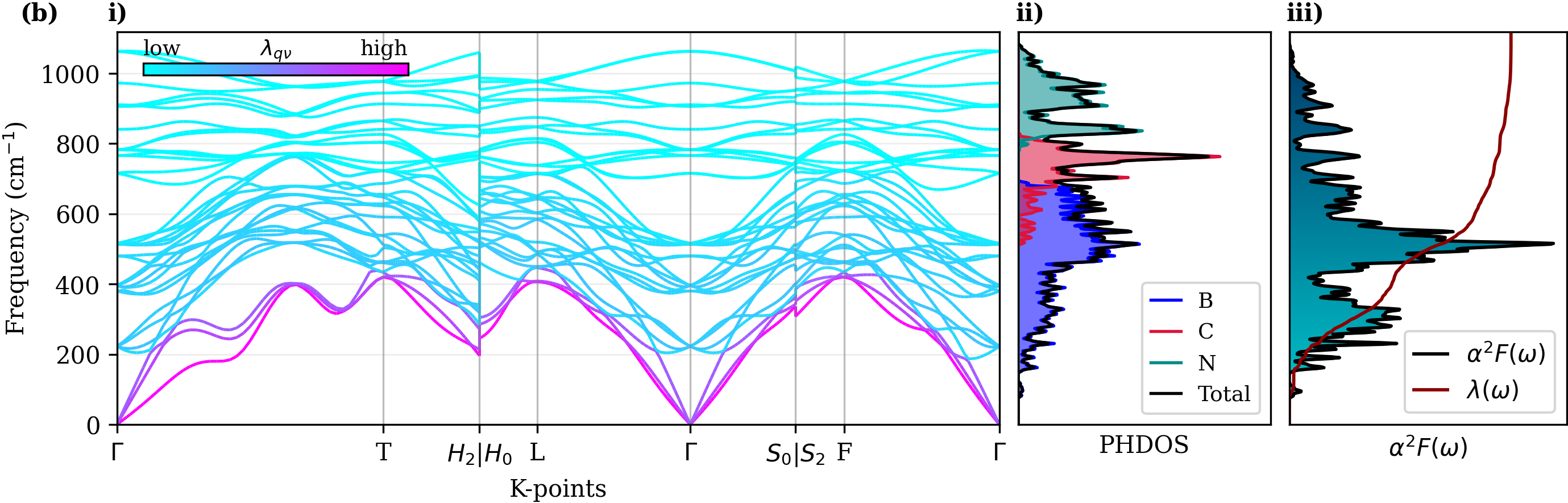}\\
  \includegraphics[width=\textwidth]{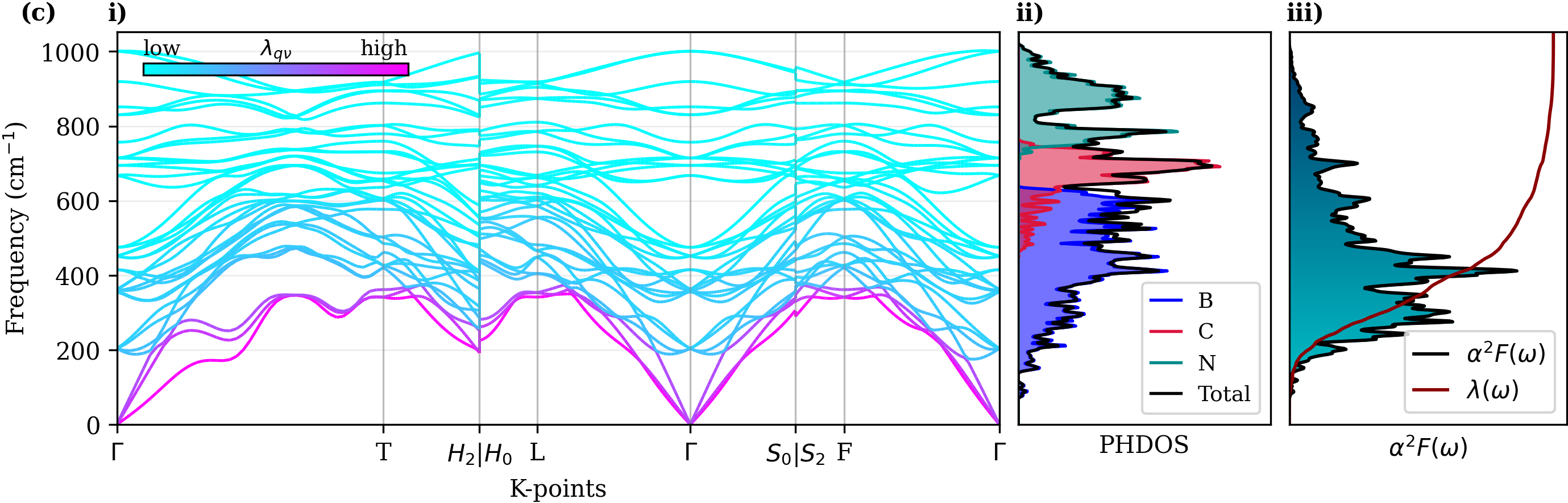}\\
  \includegraphics[width=\textwidth]{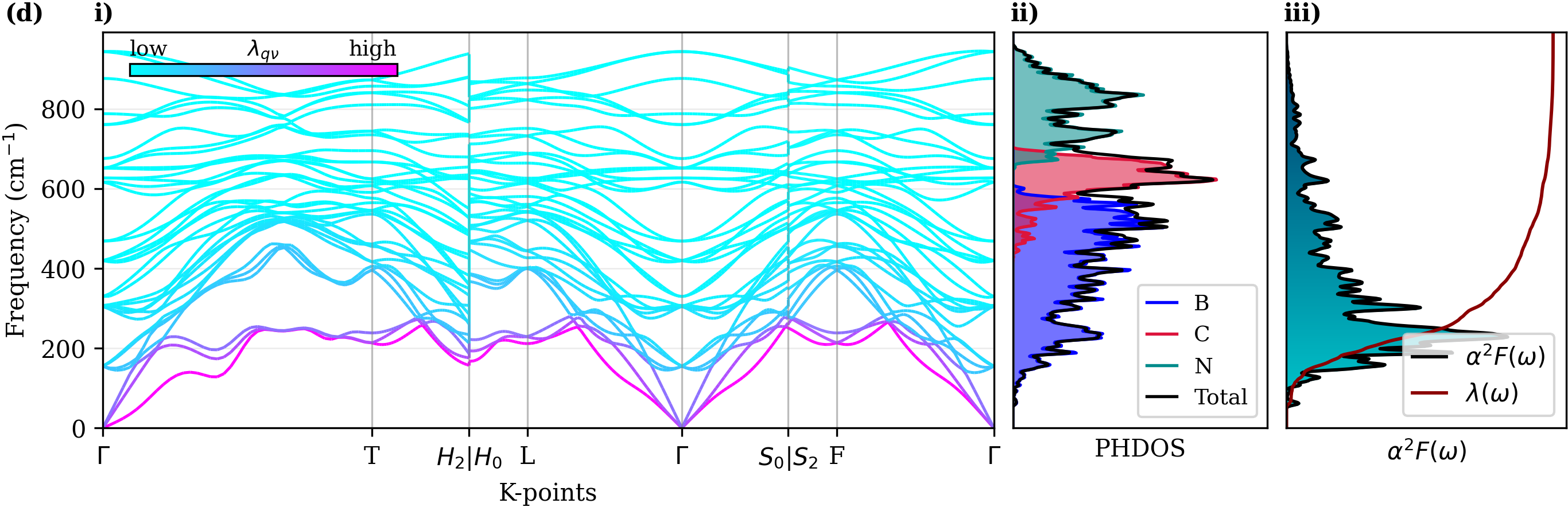}
  \caption{i) Phonon dispersions along the high-symmetry path, ii) projected phonon density of states (PHDOS), and iii) Eliashberg spectral function $\alpha^2F(\omega)$ with cumulative coupling $\lambda(\omega)$ for R3m B$_2$CN under (a) 0\%, (b) 2\%, (c) 4\%, and (d) 6\% isotropic tensile strain. Branch colors in (i) indicate  mode-resolved coupling strength $\lambda_{q\nu}$ from low (cyan) to high (magenta).}
  \label{fig:phonon}
\end{figure*}

Phonon dispersions were computed for all four strained structures along  $\Gamma$--T--H$_2|$H$_0$--L--$\Gamma$--S$_0|$--S$_2$--F--$\Gamma$    of the hexagonal Brillouin zone. No imaginary frequencies appeared at any q-point or strain level. All four configurations are dynamically stable~\cite{baroni2001phonons}.

The phonon spectrum of B$_2$CN is shown in Fig.~\ref{fig:phonon}(i). At zero strain the spectrum extends to approximately 1100~cm$^{-1}$. Atom-projected phonon density of states (PHDOS) analysis (Fig.~\ref{fig:phonon}(ii)) shows that the vibrational character is governed by interatomic force constants rather than atomic mass alone. N-derived modes dominate the high-frequency region (roughly 600-1100 cm$^{-1}$), where N sits in a stiff bonding environment in the lattice. C modes fall mainly in the mid-frequency range. B modes are concentrated at lower frequencies. This happens not because boron is the lightest atom, but because the restoring forces at B sites are comparatively softer. Phonon frequencies scale as $\omega \propto \sqrt{k_{\mathrm{eff}}/m}$, so a soft local force constant can outweigh a light atomic mass and pin the associated modes toward lower frequency~\cite{ashcroft1976solid}. This is consistent with the bond-projected force constants in Table~\ref{stab:fc}. This force-constant hierarchy remains qualitatively unchanged across the strain series, and the identity of the vibrational eigenstates is preserved even as their frequencies shift.

Isotropic tensile strain progressively softens the entire phonon spectrum~\cite{kohn1959image}. The maximum phonon frequency falls from approximately  1100~cm$^{-1}$ at 0\% to roughly 900~cm$^{-1}$ at 6\% strain. All three PHDOS envelopes shift to lower frequencies with each lattice increment (Fig.~\ref{fig:phonon}(ii)). This global softening follows from the weakening of interatomic force constants as B-C and B-N bonds elongate under lattice expansion. At 6\% strain, the lowest acoustic and low-optical branches near $\Gamma$, T \& F show a pronounced downward shift without developing imaginary character. The system is approaching but has not yet reached a lattice instability. The strain-induced softening of the B-dominated acoustic branches can be tested using inelastic X-ray or neutron scattering~\cite{baron2009phonons,pintschovius2005electron}. Measurements of the phonon dispersion at different strain levels would directly test the calculated softening and provide experimental insight into its connection with the enhanced electron--phonon coupling predicted here.

The mode-resolved coupling strength~\cite{allen1972neutron} $\lambda_{q\nu}$ is shown by the color scale from cyan (low) to magenta (high) in Fig.~\ref{fig:phonon}(iii). It is strongly concentrated in the low-to-mid frequency branches at all strain levels. At 0\% strain, the most strongly coupled modes cluster around 200--400~cm$^{-1}$, in predominantly B-character and mixed B--C branches. As strain increases, these high-$\lambda_{q\nu}$ modes shift downward — to roughly 150--250~cm$^{-1}$ at 6\% strain — driven by phonon softening in the low-frequency sector. The high-frequency N-dominated optical branches remain weakly coupled throughout. N contributes little weight at $E_F$, so these modes couple poorly to the Fermi surface.

Although no imaginary frequencies appear within the studied strain range, if the strain is increased to 8\% or 10\%, the lowest acoustic branches develop imaginary character, indicating a dynamical instability, as shown in Section~\ref{ssec:unstable_phonon}. The 6\% strain level is therefore near the upper limit of mechanical and dynamical stability for R3m B$_2$CN.

\section{Electron--Phonon Coupling \& Superconducting Critical Temperature}
\label{sec:tc}

\begin{figure*}[]
  \centering
  \includegraphics[width=\textwidth]{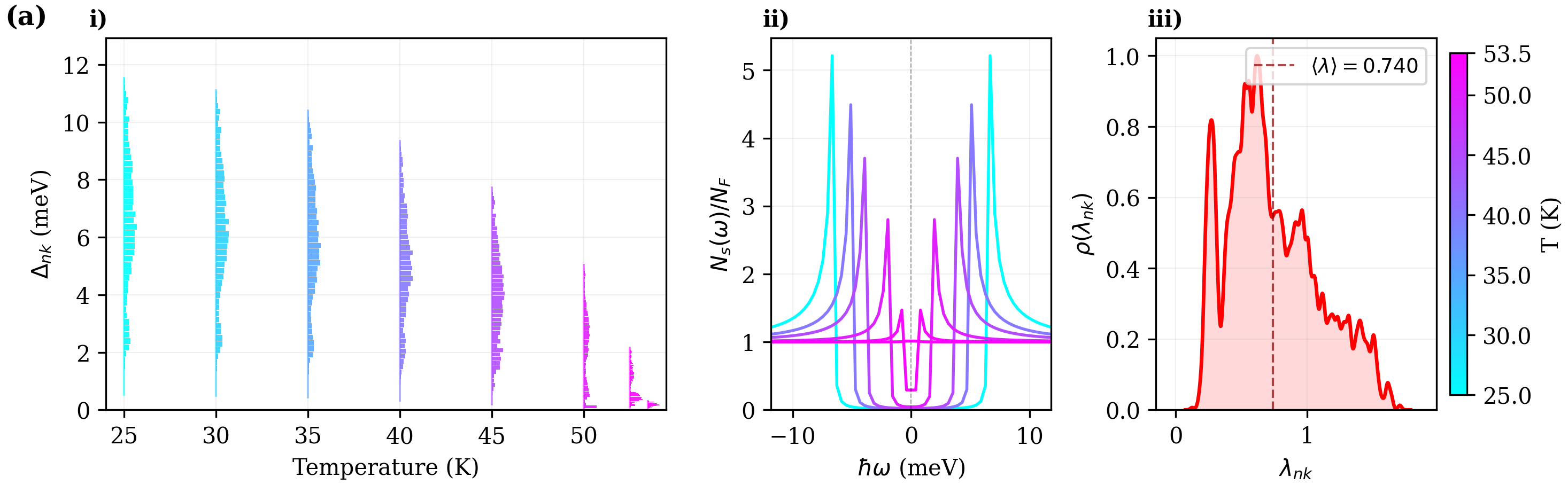}\\
  \includegraphics[width=\textwidth]{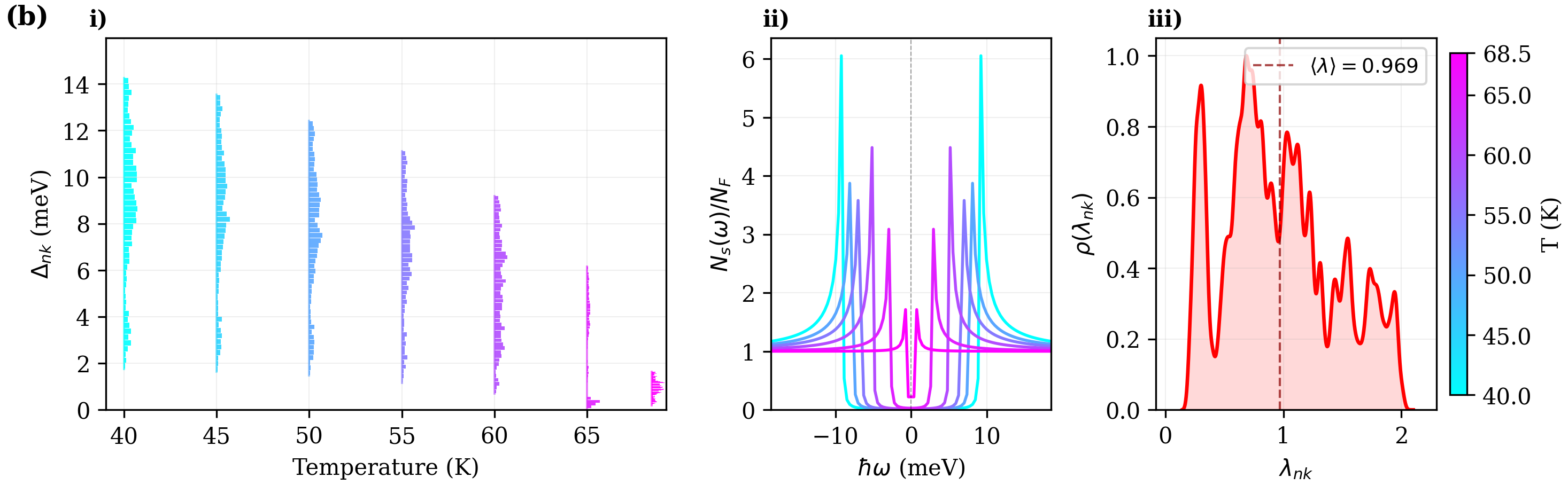}\\
  \includegraphics[width=\textwidth]{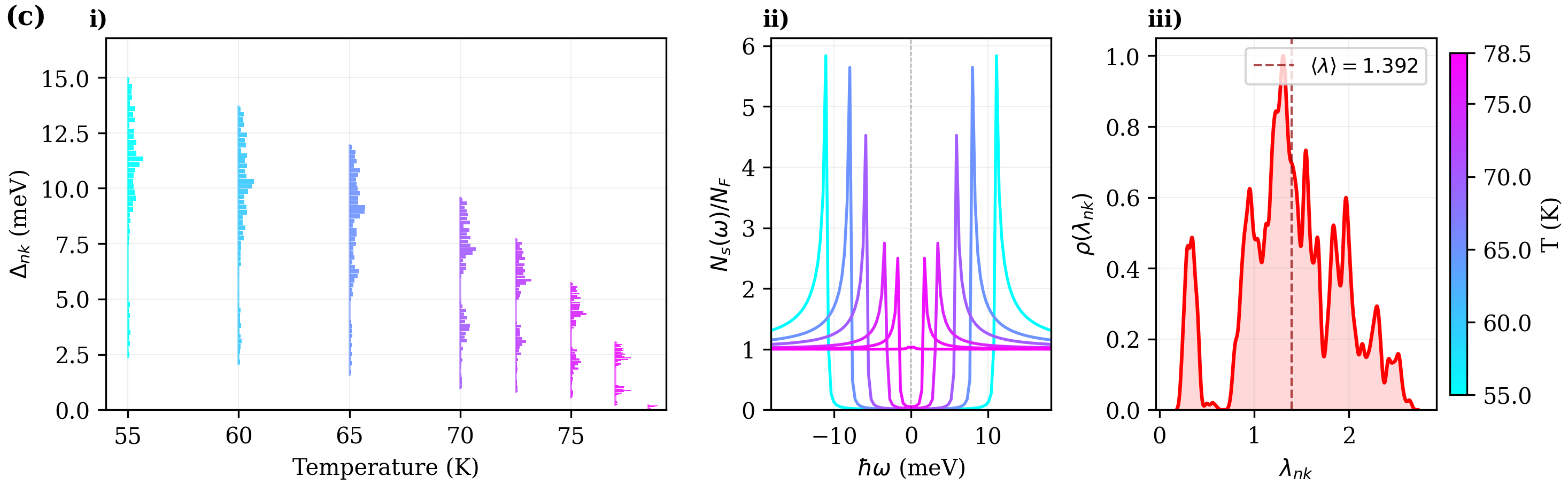}\\
  \includegraphics[width=\textwidth]{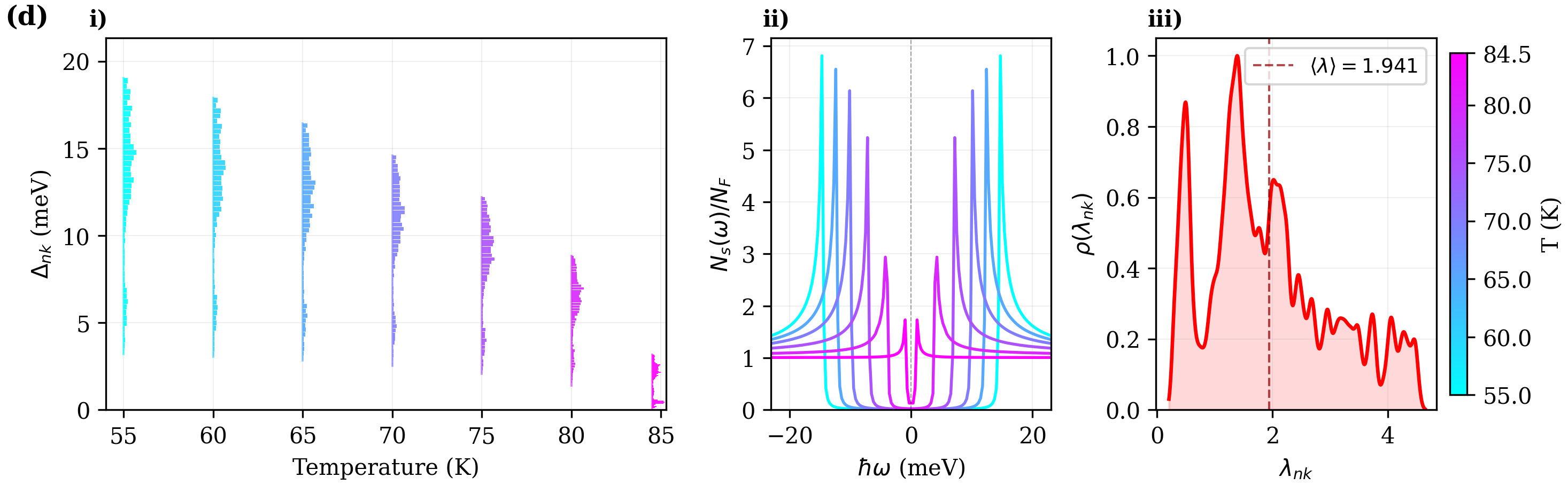}
  \caption{Anisotropic Eliashberg results for R3m B$_2$CN at (a) 0\%, (b) 2\%, (c) 4\%, and (d) 6\% isotropic tensile strain. Panel (i): anisotropic gap distribution $\Delta_{nk}$ vs. temperature (cyan: low T, magenta: near $T_c$). Panel (ii): quasiparticle DOS $N_s(\omega)/N_F$ showing coherence peaks collapsing toward $T_c$. Panel (iii): k-resolved coupling distribution $\rho(\lambda_{nk})$; dashed line marks $\langle\lambda\rangle$}
  \label{fig:aniso}
\end{figure*}

\begin{figure*}[]
  \centering
  \includegraphics[width=\textwidth]{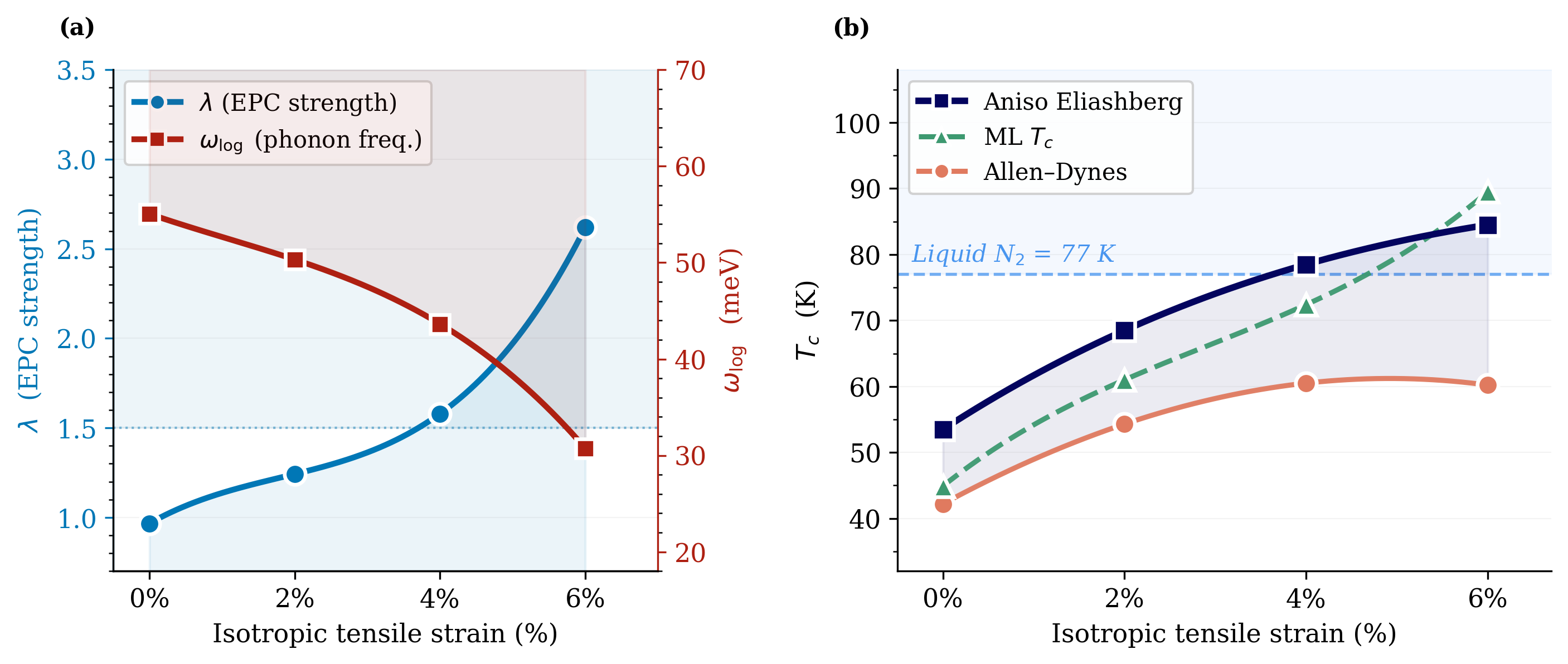}
  \caption{R3m B$_2$CN superconducting properties versus isotropic tensile strain. (a) EPC strength $\lambda$ and logarithmic average phonon frequency  $\omega_{\rm log}$. (b) Superconducting $T_c$ from the anisotropic Eliashberg equations, Allen-Dynes formula, and machine-learning predictor. Dashed line: liquid nitrogen boiling point (77 K).}
  \label{fig:summary}
\end{figure*}

\subsection{Formalism}
Sections~\ref{sec:elec} and \ref{sec:phonon} established the two inputs: band narrowing raises $N(E_F)$ and tensile strain softens the phonon spectrum. This section quantifies how they combine to determine $\lambda$ and $T_c$.

The isotropic reference point is set by the Eliashberg spectral function, 
\begin{equation}
	\alpha^{2}F(\omega)=\frac{1}{2\pi N(E_F)}
	\sum_{\mathbf{q}\nu}
	\frac{\gamma_{\mathbf{q}\nu}}
	{\omega_{\mathbf{q}\nu}}
	\delta(\omega-\omega_{\mathbf{q}\nu}),
\end{equation}
and its energy-weighted integral, 
\begin{equation}
	\lambda=2\int_{0}^{\infty}\frac{\alpha^{2}F(\omega)}{\omega}\ d\omega
\end{equation}

Both were extracted from the EPW calculations following Section~\ref{sec:method}. The logarithmic average phonon frequency, 
\begin{equation}
	\omega_{\rm log}=\exp\left[\frac{2}{\lambda}\int_{0}^{\infty}
	\frac{\alpha^{2}F(\omega)\ln\omega}{\omega} \ d\omega\right]
\end{equation}
enters the Allen-Dynes formula,
\begin{equation}
T_c^{\rm AD} = \frac{\omega_{\log}}{1.2}\exp\left[
-\frac{1.04(1+\lambda)}{\lambda-\mu^{*}(1+0.62\lambda)}\right].
\end{equation}

Equation (5) is an isotropic reference point. The $T_c$ values reported in Section~\ref{subsec:ADvsAE} come from fully anisotropic Migdal-Eliashberg equations solved on the Matsubara frequency axis in EPW.

\subsection{Electron-Phonon Coupling Strength}
The Eliashberg spectral function~\cite{allen1972neutron} $\alpha^2F(\omega)$ and its cumulative integral~\cite{dynes1972mcmillan} $\lambda(\omega)$ (Fig.~\ref{fig:phonon}, iii) track how electron-phonon coupling evolves with isotropic tensile strain. At 0\% strain, spectral weight in $\alpha^2F(\omega)$ is distributed broadly across low and mid frequencies, giving a coupling constant $\lambda = 0.967$. As strain increases, a pronounced low-frequency peak develops in $\alpha^2F(\omega)$ and grows. By 6\% strain, this peak dominates the spectral function, and $\lambda$ reaches 2.620, which is 2.7 times its unstrained value. The key electron-phonon and superconducting parameters for all four strain levels are collected in Table~\ref{tab:tc_results}.

\begin{table*}[t]
\setlength{\tabcolsep}{8pt}
\centering
\begin{tabular}{ccccccc}
\hline
Strain & $\lambda$ & $\omega_{\rm log}$ & BCS $\Delta_0$ & Allen--Dynes $T_c$ & ML $T_c$ & Aniso $T_c$ \\
\hline
0\% & 0.967 & 55.04 & 6.40 & 42.2 K & 44.9 K & 53.5 K \\
2\% & 1.242 & 50.32 & 8.27 & 54.5 K & 61.0 K & 68.5 K \\
4\% & 1.579 & 43.64 & 9.18 & 60.5 K & 72.4 K & 78.5 K \\
6\% & 2.620 & 30.74 & 9.13 & 60.2 K & 89.5 K & 84.5 K \\
\hline
\end{tabular}
\caption{Superconducting parameters and critical temperatures of R3m B$_2$CN under different isotropic tensile strain, showing the electron--phonon coupling constant $\lambda$, the logarithmic average phonon frequency $\omega_{\rm log}$ (in meV), the superconducting gap at zero temperature $\Delta_0$ (in meV), and the superconducting transition temperatures $T_c$ calculated using the Allen--Dynes equation, a machine-learning (ML) model, and an anisotropic Eliashberg calculation (Aniso), respectively.}
\label{tab:tc_results}
\end{table*}

The cumulative $\lambda(\omega)$ curves make the regime change concrete. At 0\% strain, $\lambda(\omega)$ rises gradually across the full frequency range, saturating near 1000~cm$^{-1}$. By 6\% strain it rises steeply below $\approx 300$~cm$^{-1}$ and then flattens --- more than 80\% of the total $\lambda = 2.620$ accumulates in this low-frequency window. This is a strong-coupling limit, and it has direct consequences for the Allen--Dynes approximation discussed below.

The superlinear growth of $\lambda$ with strain compounds two mechanisms acting together. Band narrowing raises $N(E_F)$ (Section~\ref{sec:elec}). At the same time, softening of the most strongly coupled low-frequency phonon branches (Section~\ref{sec:phonon}) amplifies the Fermi-surface-averaged electron--phonon matrix elements. Their combined impact is sharpest between 4\% and 6\% strain, where $\lambda$ increases by 1.04 --- more than the total gain over the entire 0--4\% range.

\subsection{Superconducting $T_c$: Allen-Dynes vs. Anisotropic Eliashberg}
\label{subsec:ADvsAE}

The strain dependence of the superconducting properties is summarized in Fig.~\ref{fig:summary}. The phonon softening that drives this enhancement simultaneously reduces $\omega_{\rm log}$ as spectral weight shifts to lower frequencies. This competition produces the non-monotonic Allen--Dynes $T_c$~\cite{allen1975transition}. $T_c^{\rm AD}$ rises from 42.2~K at 0\% to a peak of 60.5~K at 4\% strain, then stalls at 60.2~K at 6\%. This stall happens because the sharp drop in $\omega_{\rm log}$, from 43.64 to 30.74~meV, partially offsets the gain in $\lambda$. Above $\lambda \approx 1.5$ the Allen--Dynes formula loses accuracy~\cite{carbotte1990properties,marsiglio1986strong,marsiglio2020eliashberg}, a threshold crossed between 2\% and 4\% strain, exactly where the $T_c^{\rm AD}$ curve begins to flatten.

The anisotropic Eliashberg equations~\cite{allen1983theory} yield a monotonically increasing $T_c$ across the strain series (Fig.~\ref{fig:summary}b). The gap between $T_c^{\rm AD}$ and $T_c^{\rm aniso}$ widens from 11.3~K at 0\% to 24.3~K at 6\%. This growing gap is a direct measure of how badly the Allen--Dynes approximation breaks down as the system moves deeper into the strong-coupling regime. The predicted transition temperatures can be experimentally tested using four-probe resistivity and AC susceptibility measurements under controlled strain or in intercalated bulk specimens~\cite{somayazulu2019evidence,gomory1997characterization}.

\subsection{Anisotropic Gap Structure and Quasiparticle DOS}
The gap distribution $\Delta_{nk}$ as a function of temperature (Fig.~\ref{fig:aniso}, i) is strongly anisotropic at all strain levels.  It spans continuously across a finite range rather than taking a single isotropic value. At 0\% strain, gaps span roughly 0--12~meV at low temperature. With increasing strain, both the width and the upper edge grow, reaching $\approx 15$~meV at 4\% and $\approx 20$~meV at 6\% strain. At high strain the distribution stratifies into distinct bands. Each band maps onto one of the seven Fermi sheets identified in Section~\ref{sec:elec}, and sheets with stronger Fermi-surface coupling develop proportionally larger gaps. The distribution collapses uniformly to zero at  $T_c$ at every strain level, confirming a single superconducting transition with no partial gap above $T_c$.

The quasiparticle density of states $N_s(\omega)/N_F$ (Fig.~\ref{fig:aniso}, ii) encodes the gap structure directly. At each strain level sharp coherence peaks appear symmetrically about $\hbar\omega = 0$ and collapse toward $N_F$ as $T \to T_c$, as expected for a superconducting DOS~\cite{bardeen1957theory}. Their energy increases with strain, consistent with the growing maximum gap. At 6\% strain the spectrum develops a multi-peak structure extending to $|\hbar\omega| \approx 20$~meV -- each peak arising from a different Fermi sheet~\cite{souma2003origin,szabo2001evidence}. This produces spectral complexity absent in a single-gap s-wave superconductor. The multi-peak quasiparticle DOS in Fig.~\ref{fig:aniso}.(ii) can be tested by low-temperature scanning tunneling spectroscopy (STS)~\cite{fischer2007scanning}. In STS, the differential conductance provides a measure of the superconducting density of states. Point-contact Andreev reflection (PCAR) gives a complementary measurement of the superconducting gap and can also separate multiple gap scales in multiband superconductors~\cite{daghero2010probing}.

The k-resolved coupling distribution $\rho(\lambda_{nk})$ (Fig.~\ref{fig:aniso}, iii) maps how coupling strength varies across the Fermi surface. At 0\% strain, the distribution is compact, contained almost entirely within $\lambda_{nk} < 1.5$, with Fermi-surface average $\langle\lambda\rangle = 0.740$. As the strain increases it broadens and a high-coupling tail emerges. $\langle\lambda\rangle$ rises to 0.969, 1.392, and 1.941 at 2\%, 4\%, and 6\% strain, and the tail expands beyond $\lambda_{nk} > 4$ at 6\% strain. A subset of k-states on the Fermi surface becomes extremely strongly coupled under high strain. These states drive the super-linear growth of $\lambda$ and the large anisotropic $T_c$. 

\subsection{Independent Cross-Validation}
The SISSO-based machine-learning (ML) model implemented in EPW provides a complementary estimate of $T_c$~\cite{xie2022machine,lee2023electron,stanev2018machine}. The ML $T_c$ rises monotonically from 44.9~K at 0\% to 89.5~K at 6\% strain, staying within 10~K of $T_c^{\rm aniso}$ across the full range. Agreement between two independent methods -- one from first-principles Eliashberg theory and one from a data-driven model -- validates the predicted $T_c$ values. It also confirms the monotonic trend that the Allen--Dynes formula fails to capture.

The BCS superconducting gap~\cite{bardeen1957theory} $\Delta_0$ increases from 6.40~meV at 0\% to 9.18~meV at 4\% strain, then holds at 9.13~meV at 6\% as $\lambda$ continues to rise sharply. In this strong-coupling regime the gap saturates while $\lambda$ keeps growing -- the two are no longer simply proportional. The picture across the strain levels is clear: $\lambda$ grows superlinearly, $T_c^{\rm aniso}$ increases by 31~K over the 0--6\% range.

\section{Conclusion}
Isotropic strain is a direct route to stronger superconductivity in R3m B$_2$CN. Across 0-6\% strain, $\lambda$ rises from 0.967 to 2.620 and $\omega_{\rm log}$ falls from 55.04 to 30.74~meV. The anisotropic Eliashberg $T_c$ rises monotonically from 53.5~K to 84.5~K. All expanded structures are dynamically stable, with no imaginary modes at any volume. Phonon calculations at 8\% and 10\% strain reveal imaginary modes at both volumes, placing the stability boundary just above 6\% and makes it the natural upper limit of the series. The enhancement is driven by genuine softening and redistribution of spectral weight, not by approach to a structural instability.

The Allen-Dynes and anisotropic Eliashberg predictions split above $\lambda \approx $ 1.5. $T_c^{\rm AD}$ peaks at 4\% strain (60.5~K, $\lambda = 1.579$) and stalls at 6\% (60.2~K). The anisotropic Eliashberg $T_c$ keeps rising and reaches 84.5~K at 6\%, and the ML estimate of 89.5~K follows it rather than Allen-Dynes. Agreement between these two independent methods indicates that gap anisotropy becomes significant under expansion and needs to be treated explicitly, not approximated away.

These calculations confirm a dynamically stable, first-principles route to strain-tunable, phonon-mediated high-$T_c$ superconductivity in a light-element covalent compound~\cite{moussa2008constraints}. This result holds across the full 0-6\% strain series, with no sign of structural instability at any point.

Reaching this experimentally means finding a way to expand the lattice isotropically. B$_2$CN's stiffness (bulk modulus $\approx$ 320~GPa) rules out mechanical loading, and epitaxial strain is directional~\cite{schlom2007strain}. Intercalation of spacer species between the layers - negative chemical pressure - is the natural candidate. In practice, however, intercalation typically expands the c-axis preferentially rather than isotropically. Whether c-axis-dominated expansion reproduces the coupling enhancement seen here is the immediate open question.

The isotropic protocol isolates the volumetric contribution to electron-phonon coupling from directional and ionic-relaxation effects. It is a deliberate choice that cleanly separates lattice expansion from everything else. Modeling specific intercalants and accounting for anharmonicity~\cite{errea2014anharmonic} at large expansions are beyond the scope of this work. The stability and scale of the predicted enhancement give B$_2$CN and related ternary boron-carbon-nitrogen phases a clear case for experimental follow-up. Brillouin light scattering can give the strain-dependent elastic tensor. ARPES on strained specimens can resolve the bandwidth compression and $N(E_F)$ rise. Inelastic X-ray or neutron scattering can map the full phonon softening. Four-probe resistivity and AC susceptibility can test the predicted $T_c$ values, and scanning tunneling spectroscopy or PCAR can resolve the anisotropic multi-gap structure in Fig.~\ref{fig:aniso}.(ii). Each measurement targets a different link in the coupling chain, but taken together, they form a tractable experimental test of the strain-tuning mechanism proposed here.

\section{Data availability}
The datasets generated and analysed during this study are not publicly available due to their large file sizes, but are available from the corresponding author upon reasonable request.

\bibliographystyle{ieeetr}
\bibliography{references}

@article{giustino2017electron,
  title={Electron-phonon interactions from first principles},
  author={Giustino, Feliciano},
  journal={Reviews of Modern Physics},
  volume={89},
  number={1},
  pages={015003},
  year={2017},
  publisher={APS}
}

@article{flores2020perspective,
  title={A perspective on conventional high-temperature superconductors at high pressure: Methods and materials},
  author={Flores-Livas, Jos{\'e} A and Boeri, Lilia and Sanna, Antonio and Profeta, Gianni and Arita, Ryotaro and Eremets, Mikhail},
  journal={Physics Reports},
  volume={856},
  pages={1--78},
  year={2020},
  publisher={Elsevier}
}

@article{keimer2015quantum,
  title={From quantum matter to high-temperature superconductivity in copper oxides},
  author={Keimer, Bernhard and Kivelson, Steven A and Norman, Michael R and Uchida, Shinichi and Zaanen, J},
  journal={Nature},
  volume={518},
  number={7538},
  pages={179--186},
  year={2015},
  publisher={Nature Publishing Group UK London}
}

@article{paglione2010high,
  title={High-temperature superconductivity in iron-based materials},
  author={Paglione, Johnpierre and Greene, Richard L},
  journal={Nature physics},
  volume={6},
  number={9},
  pages={645--658},
  year={2010},
  publisher={Nature Publishing Group UK London}
}

@article{wan2024prediction,
  title={Prediction of high-temperature superconductors at ambient pressure with diamond-like structures: B 2 CX (X= N, P)},
  author={Wan, Yi and Chen, Ying-Jie and Qiao, Shu-Xiang and Jiang, Kai-Yue and Liu, Guo-Hua and Jiao, Na and Zhang, Ping and Lu, Hong-Yan},
  journal={Physical Chemistry Chemical Physics},
  volume={26},
  number={45},
  pages={28643--28649},
  year={2024},
  publisher={Royal Society of Chemistry}
}

@article{calandra2008high,
  title={High-T c superconductivity in superhard diamondlike BC 5},
  author={Calandra, Matteo and Mauri, Francesco},
  journal={Physical review letters},
  volume={101},
  number={1},
  pages={016401},
  year={2008},
  publisher={APS}
}

@article{solozhenko2009ultimate,
  title={Ultimate metastable solubility of boron in diamond: synthesis of superhard diamondlike BC 5},
  author={Solozhenko, Vladimir L and Kurakevych, Oleksandr O and Andrault, Denis and Le Godec, Yann and Mezouar, Mohamed},
  journal={Physical Review Letters},
  volume={102},
  number={1},
  pages={015506},
  year={2009},
  publisher={APS}
}

@article{smith2026first,
  title={First-principles calculation of the superconducting T c in superhard BCN metals},
  author={Smith, Adam D and Vohra, Yogesh K and Chen, Cheng-Chien},
  journal={Physical Review B},
  volume={113},
  number={21},
  pages={214509},
  year={2026},
  publisher={APS}
}

@article{tomassetti2024prospect,
  title={Prospect of high-temperature superconductivity in layered metal borocarbides},
  author={Tomassetti, Charlsey R and Kafle, Gyanu P and Marcial, Edan T and Margine, Elena R and Kolmogorov, Aleksey N},
  journal={Journal of Materials Chemistry C},
  volume={12},
  number={13},
  pages={4870--4884},
  year={2024},
  publisher={The Royal Society of Chemistry}
}

@article{li2019hard,
  title={Hard BN clathrate superconductors},
  author={Li, Xue and Yong, Xue and Wu, Min and Lu, Siyu and Liu, Hanyu and Meng, Sheng and Tse, John S and Li, Yinwei},
  journal={The Journal of Physical Chemistry Letters},
  volume={10},
  number={10},
  pages={2554--2560},
  year={2019},
  publisher={ACS Publications}
}

@article{ding2024hydrogen,
  title={Hydrogen-Doped c-BN as a Promising Path to High-Temperature Superconductivity Above 120 K at Ambient Pressure},
  author={Ding, Han-Bin and Niu, Rui and Li, Shen-Ao and Liu, Ying-Ming and Chen, Xiao-Jia and Lin, Hai-Qing and Zhong, Guo-Hua},
  journal={Advanced Science},
  volume={11},
  number={44},
  pages={2408275},
  year={2024},
  publisher={Wiley Online Library}
}

@article{saha2020high,
  title={High-temperature conventional superconductivity in the boron-carbon system: Material trends},
  author={Saha, Santanu and Di Cataldo, Simone and Amsler, Maximilian and Von Der Linden, Wolfgang and Boeri, Lilia},
  journal={Physical Review B},
  volume={102},
  number={2},
  pages={024519},
  year={2020},
  publisher={APS}
}

@book{ashcroft1976solid,
  author    = {N. W. Ashcroft and N. D. Mermin},
  title     = {Solid State Physics},
  publisher = {Holt, Rinehart and Winston},
  address   = {New York},
  year      = {1976}
}

@article{bardeen1957theory,
  title={Theory of superconductivity},
  author={Bardeen, John and Cooper, Leon N and Schrieffer, John Robert},
  journal={Physical review},
  volume={108},
  number={5},
  pages={1175},
  year={1957},
  publisher={APS}
}

@article{bardeen1957microscopic,
  title={Microscopic theory of superconductivity},
  author={Bardeen, John and Cooper, Leon N and Schrieffer, J Robert},
  journal={Physical Review},
  volume={106},
  number={1},
  pages={162},
  year={1957},
  publisher={APS}
}

@article{migdal1958interaction,
  title={Interaction between electrons and lattice vibrations in a normal metal},
  author={Migdal, AB},
  journal={Sov. Phys. JETP},
  volume={7},
  number={6},
  pages={996--1001},
  year={1958}
}

@article{eliashberg1960interactions,
  title={Interactions between electrons and lattice vibrations in a superconductor},
  author={Eliashberg, GM},
  journal={Sov. Phys. JETP},
  volume={11},
  number={3},
  pages={696--702},
  year={1960}
}

@article{baroni2001phonons,
  title={Phonons and related crystal properties from density-functional perturbation theory},
  author={Baroni, Stefano and De Gironcoli, Stefano and Dal Corso, Andrea and Giannozzi, Paolo},
  journal={Reviews of modern Physics},
  volume={73},
  number={2},
  pages={515},
  year={2001},
  publisher={APS}
}

@article{giustino2007electron,
  title={Electron-phonon interaction using Wannier functions},
  author={Giustino, Feliciano and Cohen, Marvin L and Louie, Steven G},
  journal={Physical Review B—Condensed Matter and Materials Physics},
  volume={76},
  number={16},
  pages={165108},
  year={2007},
  publisher={APS}
}

@article{pickard2020superconducting,
  title={Superconducting hydrides under pressure},
  author={Pickard, Chris J and Errea, Ion and Eremets, Mikhail I},
  journal={Annual Review of Condensed Matter Physics},
  volume={11},
  number={1},
  pages={57--76},
  year={2020},
  publisher={Annual Reviews}
}

@article{cerqueira2024sampling,
  title={Sampling the materials space for conventional superconducting compounds},
  author={Cerqueira, Tiago FT and Sanna, Antonio and Marques, Miguel AL},
  journal={Advanced Materials},
  volume={36},
  number={1},
  pages={2307085},
  year={2024},
  publisher={Wiley Online Library}
}

@article{kortus2001superconductivity,
  title={Superconductivity of metallic boron in MgB 2},
  author={Kortus, Jens and Mazin, II and Belashchenko, Kirill D and Antropov, Vladimir P and Boyer, LL},
  journal={Physical Review Letters},
  volume={86},
  number={20},
  pages={4656},
  year={2001},
  publisher={APS}
}

@article{li2022superconductivity,
  title={Superconductivity above 200 K discovered in superhydrides of calcium},
  author={Li, Zhiwen and He, Xin and Zhang, Changling and Wang, Xiancheng and Zhang, Sijia and Jia, Yating and Feng, Shaomin and Lu, Ke and Zhao, Jianfa and Zhang, Jun and others},
  journal={Nature communications},
  volume={13},
  number={1},
  pages={2863},
  year={2022},
  publisher={Nature Publishing Group UK London}
}

@article{ma2022high,
  title={High-temperature superconducting phase in clathrate calcium hydride CaH 6 up to 215 K at a pressure of 172 GPa},
  author={Ma, Liang and Wang, Kui and Xie, Yu and Yang, Xin and Wang, Yingying and Zhou, Mi and Liu, Hanyu and Yu, Xiaohui and Zhao, Yongsheng and Wang, Hongbo and others},
  journal={Physical Review Letters},
  volume={128},
  number={16},
  pages={167001},
  year={2022},
  publisher={APS}
}

@article{sanna2024prediction,
  title={Prediction of ambient pressure conventional superconductivity above 80 K in hydride compounds},
  author={Sanna, Antonio and Cerqueira, Tiago FT and Fang, Yue-Wen and Errea, Ion and Ludwig, Alfred and Marques, Miguel AL},
  journal={npj Computational Materials},
  volume={10},
  number={1},
  pages={44},
  year={2024},
  publisher={Nature Publishing Group UK London}
}

@article{gao2025maximum,
  title={The maximum T c of conventional superconductors at ambient pressure},
  author={Gao, Kun and Cerqueira, Tiago FT and Sanna, Antonio and Fang, Yue-Wen and Dangi{\'c}, \DJ or\dj e and Errea, Ion and Wang, Hai-Chen and Botti, Silvana and Marques, Miguel AL},
  journal={Nature Communications},
  volume={16},
  number={1},
  pages={8253},
  year={2025},
  publisher={Nature Publishing Group UK London}
}

@article{ashcroft1968metallic,
  title={Metallic hydrogen: a high-temperature superconductor?},
  author={Ashcroft, Neil W},
  journal={Physical Review Letters},
  volume={21},
  number={26},
  pages={1748},
  year={1968},
  publisher={APS}
}

@article{ashcroft2004hydrogen,
  title={Hydrogen dominant metallic alloys: high temperature superconductors?},
  author={Ashcroft, NW},
  journal={Physical Review Letters},
  volume={92},
  number={18},
  pages={187002},
  year={2004},
  publisher={APS}
}

@article{nagamatsu2001superconductivity,
  title={Superconductivity at 39 K in magnesium diboride},
  author={Nagamatsu, Jun and Nakagawa, Norimasa and Muranaka, Takahiro and Zenitani, Yuji and Akimitsu, Jun},
  journal={nature},
  volume={410},
  number={6824},
  pages={63--64},
  year={2001},
  publisher={Nature Publishing Group UK London}
}

@article{duan2014pressure,
  title={Pressure-induced metallization of dense (H2S) 2H2 with high-T c superconductivity},
  author={Duan, Defang and Liu, Yunxian and Tian, Fubo and Li, Da and Huang, Xiaoli and Zhao, Zhonglong and Yu, Hongyu and Liu, Bingbing and Tian, Wenjing and Cui, Tian},
  journal={Scientific reports},
  volume={4},
  number={1},
  pages={6968},
  year={2014},
  publisher={Nature Publishing Group UK London}
}

@article{drozdov2015conventional,
  title={Conventional superconductivity at 203 kelvin at high pressures in the sulfur hydride system},
  author={Drozdov, AP and Eremets, MI and Troyan, IA and Ksenofontov, Vadim and Shylin, Sergii I},
  journal={Nature},
  volume={525},
  number={7567},
  pages={73--76},
  year={2015},
  publisher={Nature Publishing Group UK London}
}

@article{drozdov2019superconductivity,
  title={Superconductivity at 250 K in lanthanum hydride under high pressures},
  author={Drozdov, AP and Kong, PP and Minkov, VS and Besedin, SP and Kuzovnikov, MA and Mozaffari, S and Balicas, L and Balakirev, Fedor Fedorovich and Graf, DE and Prakapenka, VB and others},
  journal={Nature},
  volume={569},
  number={7757},
  pages={528--531},
  year={2019},
  publisher={Nature Publishing Group UK London}
}

@article{liu1989atomic,
  title={Atomic arrangement and electronic structure of BC 2 N},
  author={Liu, Amy Y and Wentzcovitch, Renata M and Cohen, Marvin L},
  journal={Physical Review B},
  volume={39},
  number={3},
  pages={1760},
  year={1989},
  publisher={APS}
}

@article{tateyama1997proposed,
  title={Proposed synthesis path for heterodiamond BC 2 N},
  author={Tateyama, Yoshitaka and Ogitsu, T and Kusakabe, K and Tsuneyuki, S and Itoh, S},
  journal={Physical Review B},
  volume={55},
  number={16},
  pages={R10161},
  year={1997},
  publisher={APS}
}

@article{solozhenko2001synthesis,
  title={Synthesis of superhard cubic BC 2 N},
  author={Solozhenko, Vladimir L and Andrault, Denis and Fiquet, Guillaume and Mezouar, Mohamed and Rubie, David C},
  journal={Applied Physics Letters},
  volume={78},
  number={10},
  pages={1385--1387},
  year={2001},
  publisher={American Institute of Physics}
}

@article{knittle1989experimental,
  title={Experimental and theoretical equation of state of cubic boron nitride},
  author={Knittle, Elise and Wentzcovitch, Renata M and Jeanloz, Raymond and Cohen, Marvin L},
  journal={Nature},
  volume={337},
  number={6205},
  pages={349--352},
  year={1989},
  publisher={Nature Publishing Group UK London}
}

@article{watanabe2004direct,
  title={Direct-bandgap properties and evidence for ultraviolet lasing of hexagonal boron nitride single crystal},
  author={Watanabe, Kenji and Taniguchi, Takashi and Kanda, Hisao},
  journal={Nature materials},
  volume={3},
  number={6},
  pages={404--409},
  year={2004},
  publisher={Nature Publishing Group UK London}
}

@article{bekaert2017evolution,
  title={Evolution of multi-gap superconductivity in the atomically thin limit: Strain-enhanced three-gap superconductivity in monolayer MgB $ \_2$},
  author={Bekaert, J and Aperis, Alex and Partoens, B and Oppeneer, Peter M and Milosevic, MV},
  journal={arXiv preprint arXiv:1708.04327},
  year={2017}
}

@article{locquet1998doubling,
  title={Doubling the critical temperature of La1. 9Sr0. 1CuO4 using epitaxial strain},
  author={Locquet, J-P and Perret, J and Fompeyrine, J and M{\"a}chler, E and Seo, Jin Won and Van Tendeloo, Gustaaf},
  journal={Nature},
  volume={394},
  number={6692},
  pages={453--456},
  year={1998},
  publisher={Nature Publishing Group}
}

@article{PhysRevLett.93.147006,
  title = {Enhancement of the Superconducting Transition Temperature of ${\mathrm{M}\mathrm{g}\mathrm{B}}_{2}$ by a Strain-Induced Bond-Stretching Mode Softening},
  author = {Pogrebnyakov, A. V. and Redwing, J. M. and Raghavan, S. and Vaithyanathan, V. and Schlom, D. G. and Xu, S. Y. and Li, Qi and Tenne, D. A. and Soukiassian, A. and Xi, X. X. and Johannes, M. D. and Kasinathan, D. and Pickett, W. E. and Wu, J. S. and Spence, J. C. H.},
  journal = {Phys. Rev. Lett.},
  volume = {93},
  issue = {14},
  pages = {147006},
  numpages = {4},
  year = {2004},
  month = {Sep},
  publisher = {American Physical Society},
  doi = {10.1103/PhysRevLett.93.147006},
  url = {https://link.aps.org/doi/10.1103/PhysRevLett.93.147006}
}

@article{henriquez2024modulation,
  title={Modulation of the superconducting phase transition in multilayer 2H-NbSe2 induced by uniform biaxial compressive strain},
  author={Henr{\'\i}quez-Guerra, Eudomar and Almonte, Lisa and Li, Hao and Elvira, Daniel and Calvo, M Reyes and Castellanos-Gomez, Andres},
  journal={Nano letters},
  volume={24},
  number={34},
  pages={10504--10509},
  year={2024},
  publisher={ACS Publications}
}

@article{johansson2022effect,
  title={The effect of strain and pressure on the electron-phonon coupling and superconductivity in MgB2—Benchmark of theoretical methodologies and outlook for nanostructure design},
  author={Johansson, Erik and Tasn{\'a}di, Ferenc and Ektarawong, Annop and Rosen, Johanna and Alling, Bj{\"o}rn},
  journal={Journal of Applied Physics},
  volume={131},
  number={6},
  year={2022},
  publisher={AIP Publishing}
}

@article{kohn1965self,
  title={Self-consistent equations including exchange and correlation effects},
  author={Kohn, Walter and Sham, Lu Jeu},
  journal={Physical review},
  volume={140},
  number={4A},
  pages={A1133},
  year={1965},
  publisher={APS}
}

@article{hohenberg1964inhomogeneous,
  title={Inhomogeneous electron gas},
  author={Hohenberg, Pierre and Kohn, Walter},
  journal={Physical review},
  volume={136},
  number={3B},
  pages={B864},
  year={1964},
  publisher={APS}
}

@article{giannozzi2009quantum,
  title={QUANTUM ESPRESSO: a modular and open-source software project for quantum simulations of materials},
  author={Giannozzi, Paolo and Baroni, Stefano and Bonini, Nicola and Calandra, Matteo and Car, Roberto and Cavazzoni, Carlo and Ceresoli, Davide and Chiarotti, Guido L and Cococcioni, Matteo and Dabo, Ismaila and others},
  journal={Journal of physics: Condensed matter},
  volume={21},
  number={39},
  pages={395502},
  year={2009}
}

@article{giannozzi2017advanced,
  title={Advanced capabilities for materials modelling with Quantum ESPRESSO},
  author={Giannozzi, Paolo and Andreussi, Oliviero and Brumme, Thomas and Bunau, Oana and Buongiorno Nardelli, M and Calandra, Matteo and Car, Roberto and Cavazzoni, Carlo and Ceresoli, Davide and Cococcioni, Matteo and others},
  journal={Journal of physics: Condensed matter},
  volume={29},
  number={46},
  pages={465901},
  year={2017},
  publisher={IOP Publishing}
}

@article{giannozzi2020quantum,
  title={Quantum ESPRESSO toward the exascale},
  author={Giannozzi, Paolo and Baseggio, Oscar and Bonf{\`a}, Pietro and Brunato, Davide and Car, Roberto and Carnimeo, Ivan and Cavazzoni, Carlo and De Gironcoli, Stefano and Delugas, Pietro and Ferrari Ruffino, Fabrizio and others},
  journal={The Journal of chemical physics},
  volume={152},
  number={15},
  year={2020},
  publisher={AIP Publishing}
}

@article{vanderbilt1990soft,
  title={Soft self-consistent pseudopotentials in a generalized eigenvalue formalism},
  author={Vanderbilt, David},
  journal={Physical review B},
  volume={41},
  number={11},
  pages={7892},
  year={1990},
  publisher={APS}
}

@article{perdew1996generalized,
  title={Generalized gradient approximation made simple},
  author={Perdew, John P and Burke, Kieron and Ernzerhof, Matthias},
  journal={Physical review letters},
  volume={77},
  number={18},
  pages={3865},
  year={1996},
  publisher={APS}
}

@article{prandini2018precision,
  title={Precision and efficiency in solid-state pseudopotential calculations},
  author={Prandini, Gianluca and Marrazzo, Antimo and Castelli, Ivano E and Mounet, Nicolas and Marzari, Nicola},
  journal={npj Computational Materials},
  volume={4},
  number={1},
  pages={72},
  year={2018},
  publisher={Nature Publishing Group UK London}
}

@article{elsasser1994density,
  title={Density-functional energies and forces with Gaussian-broadened fractional occupations},
  author={Els{\"a}sser, C and F{\"a}hnle, M and Chan, CT and Ho, KM},
  journal={Physical Review B},
  volume={49},
  number={19},
  pages={13975},
  year={1994},
  publisher={APS}
}

@article{monkhorst1976special,
  title={Special points for Brillouin-zone integrations},
  author={Monkhorst, Hendrik J and Pack, James D},
  journal={Physical review B},
  volume={13},
  number={12},
  pages={5188},
  year={1976},
  publisher={APS}
}

@article{blochl1994improved,
  title={Improved tetrahedron method for Brillouin-zone integrations},
  author={Bl{\"o}chl, Peter E and Jepsen, Ove and Andersen, Ole Krogh},
  journal={Physical Review B},
  volume={49},
  number={23},
  pages={16223},
  year={1994},
  publisher={APS}
}

@article{kawamura2019fermisurfer,
  title={FermiSurfer: Fermi-surface viewer providing multiple representation schemes},
  author={Kawamura, Mitsuaki},
  journal={Computer Physics Communications},
  volume={239},
  pages={197--203},
  year={2019},
  publisher={Elsevier}
}

@article{murnaghan1944compressibility,
  title={The compressibility of media under extreme pressures},
  author={Murnaghan, Francis Dominic},
  journal={Proceedings of the National Academy of Sciences},
  volume={30},
  number={9},
  pages={244--247},
  year={1944}
}

@article{birch1947finite,
  title={Finite elastic strain of cubic crystals},
  author={Birch, Francis},
  journal={Physical review},
  volume={71},
  number={11},
  pages={809},
  year={1947},
  publisher={APS}
}

@article{wu2005systematic,
  title={Systematic treatment of displacements, strains, and electric fields in density-functional perturbation theory},
  author={Wu, Xifan and Vanderbilt, David and Hamann, DR},
  journal={Physical Review B—Condensed Matter and Materials Physics},
  volume={72},
  number={3},
  pages={035105},
  year={2005},
  publisher={APS}
}

@misc{thermo_pw,
  author       = {Dal Corso, Andrea},
  title        = {{Thermo\_pw}},
  howpublished = {\url{https://dalcorso.github.io/thermo_pw/}},
  note         = {Accessed: 2026-07-24},
  year         = {2024}
}

@book{born1996dynamical,
  title={Dynamical theory of crystal lattices},
  author={Born, Max and Huang, Kun},
  year={1996},
  publisher={Oxford university press}
}

@article{gaillac2016elate,
  title={ELATE: an open-source online application for analysis and visualization of elastic tensors},
  author={Gaillac, Romain and Pullumbi, Pluton and Coudert, Fran{\c{c}}ois-Xavier},
  journal={Journal of Physics: Condensed Matter},
  volume={28},
  number={27},
  pages={275201},
  year={2016},
  publisher={IOP Publishing}
}

@article{ponce2016epw,
  title={EPW: Electron--phonon coupling, transport and superconducting properties using maximally localized Wannier functions},
  author={Ponc{\'e}, Samuel and Margine, Elena R and Verdi, Carla and Giustino, Feliciano},
  journal={Computer Physics Communications},
  volume={209},
  pages={116--133},
  year={2016},
  publisher={Elsevier}
}

@article{noffsinger2010epw,
  title={EPW: A program for calculating the electron--phonon coupling using maximally localized Wannier functions},
  author={Noffsinger, Jesse and Giustino, Feliciano and Malone, Brad D and Park, Cheol-Hwan and Louie, Steven G and Cohen, Marvin L},
  journal={Computer Physics Communications},
  volume={181},
  number={12},
  pages={2140--2148},
  year={2010},
  publisher={Elsevier}
}

@article{lee2023electron,
  title={Electron--phonon physics from first principles using the EPW code},
  author={Lee, Hyungjun and Ponc{\'e}, Samuel and Bushick, Kyle and Hajinazar, Samad and Lafuente-Bartolome, Jon and Leveillee, Joshua and Lian, Chao and Lihm, Jae-Mo and Macheda, Francesco and Mori, Hitoshi and others},
  journal={npj Computational Materials},
  volume={9},
  number={1},
  pages={156},
  year={2023},
  publisher={Nature Publishing Group UK London}
}

@article{marzari1997maximally,
  title={Maximally localized generalized Wannier functions for composite energy bands},
  author={Marzari, Nicola and Vanderbilt, David},
  journal={Physical review B},
  volume={56},
  number={20},
  pages={12847},
  year={1997},
  publisher={APS}
}

@article{souza2001maximally,
  title={Maximally localized Wannier functions for entangled energy bands},
  author={Souza, Ivo and Marzari, Nicola and Vanderbilt, David},
  journal={Physical Review B},
  volume={65},
  number={3},
  pages={035109},
  year={2001},
  publisher={APS}
}

@article{pizzi2020wannier90,
  title={Wannier90 as a community code: new features and applications},
  author={Pizzi, Giovanni and Vitale, Valerio and Arita, Ryotaro and Bl{\"u}gel, Stefan and Freimuth, Frank and G{\'e}ranton, Guillaume and Gibertini, Marco and Gresch, Dominik and Johnson, Charles and Koretsune, Takashi and others},
  journal={Journal of Physics: Condensed Matter},
  volume={32},
  number={16},
  pages={165902},
  year={2020},
  publisher={IOP Publishing}
}

@article{morel1962calculation,
  title={Calculation of the superconducting state parameters with retarded electron-phonon interaction},
  author={Morel, P and Anderson, PW},
  journal={Physical Review},
  volume={125},
  number={4},
  pages={1263},
  year={1962},
  publisher={APS}
}

@article{hill1952elastic,
  title={The elastic behaviour of a crystalline aggregate},
  author={Hill, Richard},
  journal={Proceedings of the Physical Society. Section A},
  volume={65},
  number={5},
  pages={349--354},
  year={1952}
}

@article{anderson1963simplified,
  title={A simplified method for calculating the Debye temperature from elastic constants},
  author={Anderson, Orson L},
  journal={Journal of Physics and Chemistry of Solids},
  volume={24},
  number={7},
  pages={909--917},
  year={1963},
  publisher={Elsevier}
}

@article{kohn1959image,
  title={Image of the Fermi surface in the vibration spectrum of a metal},
  author={Kohn, Walter},
  journal={Physical Review Letters},
  volume={2},
  number={9},
  pages={393},
  year={1959},
  publisher={APS}
}

@article{baron2009phonons,
  title={Phonons in crystals using inelastic x-ray scattering},
  author={Baron, Alfred QR},
  journal={arXiv preprint arXiv:0910.5764},
  year={2009}
}

@article{pintschovius2005electron,
  title={Electron--phonon coupling effects explored by inelastic neutron scattering},
  author={Pintschovius, Lothar},
  journal={physica status solidi (b)},
  volume={242},
  number={1},
  pages={30--50},
  year={2005},
  publisher={Wiley Online Library}
}

@article{allen1972neutron,
  title={Neutron spectroscopy of superconductors},
  author={Allen, Philip B},
  journal={Physical Review B},
  volume={6},
  number={7},
  pages={2577},
  year={1972},
  publisher={APS}
}

@article{dynes1972mcmillan,
  title={McMillan's equation and the Tc of superconductors},
  author={Dynes, RC},
  journal={Solid State Communications},
  volume={10},
  number={7},
  pages={615--618},
  year={1972},
  publisher={Elsevier}
}

@article{allen1975transition,
  title={Transition temperature of strong-coupled superconductors reanalyzed},
  author={Allen, Ph B and Dynes, RC},
  journal={Physical Review B},
  volume={12},
  number={3},
  pages={905},
  year={1975},
  publisher={APS}
}

@article{margine2013anisotropic,
  title={Anisotropic migdal-eliashberg theory using wannier functions},
  author={Margine, Elena Roxana and Giustino, Feliciano},
  journal={Physical Review B—Condensed Matter and Materials Physics},
  volume={87},
  number={2},
  pages={024505},
  year={2013},
  publisher={APS}
}

@article{mouhat2014necessary,
  title={Necessary and sufficient elastic stability conditions in various crystal systems},
  author={Mouhat, F{\'e}lix and Coudert, Fran{\c{c}}ois-Xavier},
  journal={Physical review B},
  volume={90},
  number={22},
  pages={224104},
  year={2014},
  publisher={APS}
}

@article{carlotti1997comparative,
  title={Comparative study of the elastic properties of polycrystalline aluminum nitride films on silicon by Brillouin light scattering},
  author={Carlotti, Giovanni and Gubbiotti, Gianluca and Hickernell, FS and Liaw, HM and Socino, Giovanni},
  journal={Thin Solid Films},
  volume={310},
  number={1-2},
  pages={34--38},
  year={1997},
  publisher={Elsevier}
}

@article{carlotti2005elastic,
  title={Elastic constants of low-k and barrier dielectric films measured by Brillouin light scattering},
  author={Carlotti, Giovanni and Cherault, N and Casanova, N and Goldberg, C and Socino, Giovanni},
  journal={Thin Solid Films},
  volume={493},
  number={1-2},
  pages={175--178},
  year={2005},
  publisher={Elsevier}
}

@article{clerc1998mechanical,
  title={Mechanical hardness: A semiempirical theory based on screened electrostatics and elastic shear},
  author={Clerc, DG and Ledbetter, HM},
  journal={Journal of Physics and Chemistry of Solids},
  volume={59},
  number={6-7},
  pages={1071--1095},
  year={1998},
  publisher={Elsevier}
}

@article{kral1998critical,
  title={Critical review on the elastic properties of transition metal carbides, nitrides and carbonitrides},
  author={Kral, Christoph and Lengauer, Walter and Rafaja, David and Ettmayer, Peter},
  journal={Journal of Alloys and Compounds},
  volume={265},
  number={1-2},
  pages={215--233},
  year={1998},
  publisher={Elsevier}
}

@article{kastuar2023mechanical,
  title={Mechanical properties of cubic boron nitride and diamond at dynamical pressure and temperature},
  author={Kastuar, Srihari M and Liu, Zhong-Li and Najmaei, Sina and Ekuma, Chinedu E},
  journal={Applied Physics Letters},
  volume={123},
  number={23},
  year={2023},
  publisher={AIP Publishing}
}

@article{klein1993young,
  title={Young's modulus and Poisson's ratio of CVD diamond},
  author={Klein, Claude A and Cardinale, Gregory F},
  journal={Diamond and Related Materials},
  volume={2},
  number={5-7},
  pages={918--923},
  year={1993},
  publisher={Elsevier}
}

@article{pugh1954xcii,
  title={XCII. Relations between the elastic moduli and the plastic properties of polycrystalline pure metals},
  author={Pugh, SF},
  journal={The London, Edinburgh, and Dublin Philosophical Magazine and Journal of Science},
  volume={45},
  number={367},
  pages={823--843},
  year={1954},
  publisher={Taylor \& Francis}
}

@article{chen2011modeling,
  title={Modeling hardness of polycrystalline materials and bulk metallic glasses},
  author={Chen, Xing-Qiu and Niu, Haiyang and Li, Dianzhong and Li, Yiyi},
  journal={Intermetallics},
  volume={19},
  number={9},
  pages={1275--1281},
  year={2011},
  publisher={Elsevier}
}

@article{esterlis2018bound,
  title={A bound on the superconducting transition temperature},
  author={Esterlis, I and Kivelson, SA and Scalapino, DJ},
  journal={npj Quantum Materials},
  volume={3},
  number={1},
  pages={59},
  year={2018},
  publisher={Nature Publishing Group UK London}
}

@article{smith2023machine,
  title={Machine learning the relationship between Debye temperature and superconducting transition temperature},
  author={Smith, Adam D and Harris, Sumner B and Camata, Renato P and Yan, Da and Chen, Cheng-Chien},
  journal={Physical Review B},
  volume={108},
  number={17},
  pages={174514},
  year={2023},
  publisher={APS}
}

@article{grimsditch1994elastic,
  title={Elastic constants of boron nitride},
  author={Grimsditch, M and Zouboulis, ES and Polian, A},
  journal={Journal of applied physics},
  volume={76},
  number={2},
  pages={832--834},
  year={1994},
  publisher={American Institute of Physics}
}

@article{bud2001boron,
  title={Boron isotope effect in superconducting MgB 2},
  author={Bud'ko, Sergey L and Lapertot, G and Petrovic, C and Cunningham, CE and Anderson, N and Canfield, PC},
  journal={Physical Review Letters},
  volume={86},
  number={9},
  pages={1877},
  year={2001},
  publisher={APS}
}

@article{pauling1932nature,
  title={The nature of the chemical bond. IV. The energy of single bonds and the relative electronegativity of atoms},
  author={Pauling, Linus},
  journal={Journal of the American Chemical Society},
  volume={54},
  number={9},
  pages={3570--3582},
  year={1932},
  publisher={ACS Publications}
}

@article{mott1968metal,
  title={Metal-insulator transition},
  author={Mott, Nevill F},
  journal={Reviews of Modern Physics},
  volume={40},
  number={4},
  pages={677},
  year={1968},
  publisher={APS}
}

@article{lifshitz1960anomalies,
  title={Anomalies of electron characteristics of a metal in the high pressure region},
  author={Lifshitz, IM and others},
  journal={Sov. Phys. JETP},
  volume={11},
  number={5},
  pages={1130--1135},
  year={1960}
}

@article{blanter1994theory,
  title={The theory of electronic topological transitions},
  author={Blanter, Ya M and Kaganov, MI and Pantsulaya, AV and Varlamov, AA},
  journal={Physics Reports},
  volume={245},
  number={4},
  pages={159--257},
  year={1994},
  publisher={Elsevier}
}

@article{pereira2009strain,
  title={Strain engineering of graphene’s electronic structure},
  author={Pereira, Vitor M and Castro Neto, AH},
  journal={Physical review letters},
  volume={103},
  number={4},
  pages={046801},
  year={2009},
  publisher={APS}
}

@article{mcmillan1968transition,
  title={Transition temperature of strong-coupled superconductors},
  author={McMillan, WL},
  journal={Physical Review},
  volume={167},
  number={2},
  pages={331},
  year={1968},
  publisher={APS}
}

@article{hopfield1969angular,
  title={Angular momentum and transition-metal superconductivity},
  author={Hopfield, JJ},
  journal={Physical Review},
  volume={186},
  number={2},
  pages={443},
  year={1969},
  publisher={APS}
}

@article{damascelli2003angle,
  title={Angle-resolved photoemission studies of the cuprate superconductors},
  author={Damascelli, Andrea and Hussain, Zahid and Shen, Zhi-Xun},
  journal={Reviews of modern physics},
  volume={75},
  number={2},
  pages={473},
  year={2003},
  publisher={APS}
}

@article{yang2018visualizing,
  title={Visualizing electronic structures of quantum materials by angle-resolved photoemission spectroscopy},
  author={Yang, Haifeng and Liang, Aiji and Chen, Cheng and Zhang, Chaofan and Schroeter, Niels BM and Chen, Yulin},
  journal={Nature Reviews Materials},
  volume={3},
  number={9},
  pages={341--353},
  year={2018},
  publisher={Nature Publishing Group UK London}
}

@article{carbotte1990properties,
  title={Properties of boson-exchange superconductors},
  author={Carbotte, JP},
  journal={Reviews of Modern Physics},
  volume={62},
  number={4},
  pages={1027},
  year={1990},
  publisher={APS}
}

@article{allen1983theory,
  title={Theory of superconducting Tc},
  author={Allen, Philip B and Mitrovi{\'c}, Bo{\v{z}}idar},
  journal={Solid state physics},
  volume={37},
  pages={1--92},
  year={1983},
  publisher={Elsevier}
}

@article{somayazulu2019evidence,
  title={Evidence for superconductivity above 260 K in lanthanum superhydride at megabar pressures},
  author={Somayazulu, Maddury and Ahart, Muhtar and Mishra, Ajay K and Geballe, Zachary M and Baldini, Maria and Meng, Yue and Struzhkin, Viktor V and Hemley, Russell J},
  journal={Physical review letters},
  volume={122},
  number={2},
  pages={027001},
  year={2019},
  publisher={APS}
}

@article{gomory1997characterization,
  title={Characterization of high-temperature superconductors by AC susceptibility measurements},
  author={G{\"o}m{\"o}ry, Fedor},
  journal={Superconductor Science and Technology},
  volume={10},
  number={8},
  pages={523--542},
  year={1997}
}

@article{choi2002origin,
  title={The origin of the anomalous superconducting properties of MgB2},
  author={Choi, Hyoung Joon and Roundy, David and Sun, Hong and Cohen, Marvin L and Louie, Steven G},
  journal={Nature},
  volume={418},
  number={6899},
  pages={758--760},
  year={2002},
  publisher={Nature Publishing Group UK London}
}

@article{kong2001electron,
  title={Electron-phonon interaction in the normal and superconducting states of MgB 2},
  author={Kong, Y and Dolgov, OV and Jepsen, O and Andersen, OK},
  journal={Physical Review B},
  volume={64},
  number={2},
  pages={020501},
  year={2001},
  publisher={APS}
}

@article{xie2022machine,
  title={Machine learning of superconducting critical temperature from Eliashberg theory},
  author={Xie, SR and Quan, Y and Hire, AC and Deng, B and DeStefano, JM and Salinas, I and Shah, US and Fanfarillo, L and Lim, J and Kim, J and others},
  journal={npj Computational Materials},
  volume={8},
  number={1},
  pages={14},
  year={2022},
  publisher={Nature Publishing Group UK London}
}

@article{souma2003origin,
  title={The origin of multiple superconducting gaps in MgB2},
  author={Souma, S and Machida, Y and Sato, T and Takahashi, T and Matsui, H and Wang, S-C and Ding, H and Kaminski, A and Campuzano, JC and Sasaki, S and others},
  journal={Nature},
  volume={423},
  number={6935},
  pages={65--67},
  year={2003},
  publisher={Nature Publishing Group UK London}
}

@article{szabo2001evidence,
  title={Evidence for two superconducting energy gaps in MgB 2 by point-contact spectroscopy},
  author={Szab{\'o}, P and Samuely, P and Ka{\v{c}}mar{\v{c}}{\'\i}k, J and Klein, Thierry and Marcus, J and Fruchart, Daniel and Miraglia, Salvatore and Marcenat, C and Jansen, AGM},
  journal={Physical review letters},
  volume={87},
  number={13},
  pages={137005},
  year={2001},
  publisher={APS}
}

@article{fischer2007scanning,
  title={Scanning tunneling spectroscopy of high-temperature superconductors},
  author={Fischer, {\O}ystein and Kugler, Martin and Maggio-Aprile, Ivan and Berthod, Christophe and Renner, Christoph},
  journal={Reviews of Modern Physics},
  volume={79},
  number={1},
  pages={353--419},
  year={2007},
  publisher={APS}
}

@article{daghero2010probing,
  title={Probing multiband superconductivity by point-contact spectroscopy},
  author={Daghero, Dario and Gonnelli, RS},
  journal={Superconductor Science and Technology},
  volume={23},
  number={4},
  pages={043001},
  year={2010}
}

@article{stanev2018machine,
  title={Machine learning modeling of superconducting critical temperature},
  author={Stanev, Valentin and Oses, Corey and Kusne, A Gilad and Rodriguez, Efrain and Paglione, Johnpierre and Curtarolo, Stefano and Takeuchi, Ichiro},
  journal={npj Computational Materials},
  volume={4},
  number={1},
  pages={29},
  year={2018},
  publisher={Nature Publishing Group UK London}
}

@article{marsiglio1986strong,
  title={Strong-coupling corrections to Bardeen-Cooper-Schrieffer ratios},
  author={Marsiglio, F and Carbotte, JP},
  journal={Physical Review B},
  volume={33},
  number={9},
  pages={6141},
  year={1986},
  publisher={APS}
}

@article{marsiglio2020eliashberg,
  title={Eliashberg theory: A short review},
  author={Marsiglio, F},
  journal={Annals of Physics},
  volume={417},
  pages={168102},
  year={2020},
  publisher={Elsevier}
}

@article{schlom2007strain,
  title={Strain tuning of ferroelectric thin films},
  author={Schlom, Darrell G and Chen, Long-Qing and Eom, Chang-Beom and Rabe, Karin M and Streiffer, Stephen K and Triscone, Jean-Marc},
  journal={Annu. Rev. Mater. Res.},
  volume={37},
  number={1},
  pages={589--626},
  year={2007},
  publisher={Annual Reviews}
}

@article{errea2014anharmonic,
  title={Anharmonic free energies and phonon dispersions from the stochastic self-consistent harmonic approximation: Application to platinum and palladium hydrides},
  author={Errea, Ion and Calandra, Matteo and Mauri, Francesco},
  journal={Physical Review B},
  volume={89},
  number={6},
  pages={064302},
  year={2014},
  publisher={APS}
}

@article{moussa2008constraints,
  title={Constraints on T c for superconductivity in heavily boron-doped diamond},
  author={Moussa, Jonathan E and Cohen, Marvin L},
  journal={Physical Review B—Condensed Matter and Materials Physics},
  volume={77},
  number={6},
  pages={064518},
  year={2008},
  publisher={APS}
}

\end{document}


\title{Supplemental Material for:\\
Strain-Tunable High-$T_c$ Superconductivity in Layered R$\bar{3}$m B$_2$CN}

\author{Durjay Dhali, Arafat Rahman, Tareq Mahmud, and Alamgir Kabir}
\affiliation{Department of Physics, University of Dhaka, Dhaka 1000, Bangladesh}

\maketitle

\section{Computational Convergence Tests}
\label{ssec:convergence}

The kinetic-energy cutoff of 60~Ry and charge-density cutoff of 480~Ry are the values recommended for the ultrasoft pseudopotentials (USPP) in the Standard Solid-State Pseudopotentials (SSSP) precision library~\cite{prandini2018precision}. These are benchmarked against all-electron references across a wide range of crystal structures and were used directly without additional system-specific testing.

\subsection{k-Mesh Convergence}
\label{sssec:kmesh}

Table~\ref{stab:kmesh} lists $N(E_F)$ computed with Gaussian smearing ($\sigma = 0.02 $~Ry) at three k-mesh densities across all four strain levels. The monotonic decrease with mesh density reflects shrinking smearing-induced bias rather than a lack of convergence. $N(E_F)$ shifts by less than 3\% between the $24\times24\times2$ and $32\times32\times2$ meshes at all strain levels, confirming the production mesh is converged. Main-text values were obtained with the tetrahedron method on the $24\times24\times2$ mesh, giving slightly different absolute numbers but the same convergence behavior.

\begin{table} [!ht]
\setlength{\tabcolsep}{10pt}
\centering
\begin{tabular}{lcccc}
\toprule
k-mesh & $\varepsilon=0\%$ & $\varepsilon=2\%$ & $\varepsilon=4\%$ & $\varepsilon=6\%$ \\
\midrule
$16\times16\times2$   & 2.40 & 2.53 & 2.67 & 2.80 \\
$24\times24\times2$ & 2.33 & 2.42 & 2.52 & 2.63 \\
$32\times32\times2$ & 2.26 & 2.38 & 2.50 & 2.63 \\
\bottomrule
\end{tabular}
\caption{$N(E_F)$ (states eV$^{-1}$) evaluated with Gaussian smearing ($\sigma = 0.02$~Ry) at three k-mesh densities for all four strain levels. The selected production mesh is $24\times24\times2$.}
\label{stab:kmesh}
\end{table}

\section{Wannier Function Construction and Interpolation Accuracy}
\label{ssec:wannier}

\subsection{Wannier Spread and Localization}
\label{sssec:spread}

Wannierization used initial projections onto B:$2p$, C:$2p$, and N:$2p$ orbitals, giving 36 maximally localized Wannier functions per unit cell (12 atoms $\times$ 3 $p$-orbitals). Table~\ref{stab:wannier_spread} reports the total spread decomposed into $\Omega_I$, $\Omega_D$, and $\Omega_{OD}$ at each strain level.

\begin{table} [!ht]
\setlength{\tabcolsep}{10pt}
\centering
\begin{tabular}{lcccc}
\toprule
Quantity & $\varepsilon=0\%$ & $\varepsilon=2\%$ & $\varepsilon=4\%$ & $\varepsilon=6\%$ \\
\midrule
$\Omega_I$       & 38.2295 & 33.7314 & 42.6777 & 49.2011 \\
$\Omega_D$       &  3.8236 & 37.0492 &  0.2526 &  0.4162 \\
$\Omega_{OD}$    &  6.7553 &  5.6845 &  6.0551 &  8.0404 \\
$\Omega_{Total}$ & 48.8084 & 76.4651 & 48.9855 & 57.6577 \\
\bottomrule
\end{tabular}
\caption{Wannier spread components (\AA$^2$) for B:$2p$, C:$2p$, and N:$2p$ projections at each strain level. $\Omega_I$ is gauge-invariant; $\Omega_D$ and $\Omega_{OD}$ are gauge-dependent parts minimized during Wannierization.}
\label{stab:wannier_spread}
\end{table}

\subsection{Interpolation Accuracy}
\label{sssec:wann_interp}

Wannier interpolation quality was assessed by comparing interpolated band velocities and Fermi surface smoothness to directly computed NSCF eigenvalues. The interpolated bands reproduce the correct number of Fermi crossings and orbital character at $E_F$across all four strain levels. A direct band overlay is omitted because the DFT and Wannier structures were computed along different k-paths; the Wannier path follows the hexagonal zone convention used in EPW, while the DFT path follows the rhombohedral convention used in the main text. Interpolation accuracy is instead reflected in the physical consistency of $\alpha^2F(\omega)$ and the gap distributions, neither of which shows spurious features attributable to interpolation error.

\section{Structural Details}
\label{ssec:structure}

\subsection{Fractional Coordinates and Bond Lengths}
\label{sssec:coords}

Atomic fractional coordinates were held fixed at all strain levels, preserving R$\bar{3}$m symmetry by construction (see Sec.~II of the main text). Table~\ref{stab:bonds} lists the resulting interatomic distances. Since fractional coordinates are fixed, all bond lengths scale uniformly with $a$ and $c$. The B--C and B--N distances elongate by 6\% across the full strain range, weakening the interatomic force constants that drive phonon softening and the increase in electron-phonon coupling reported in the main text.

\begin{table} [!ht]
\setlength{\tabcolsep}{10pt}
\centering
\begin{tabular}{lcccc}
\toprule
Bond & $\varepsilon=0\%$ & $\varepsilon=2\%$ & $\varepsilon=4\%$ & $\varepsilon=6\%$ \\
\midrule
B--C & 1.1145 & 1.1368 & 1.1591 & 1.1814 \\
B--N & 1.0263 & 1.0468 & 1.0674 & 1.0879 \\
C--N & 2.3256 & 2.3721 & 2.4186 & 2.4651 \\
\bottomrule
\end{tabular}
\caption{Key interatomic distances (\AA) in R$\bar{3}$m B$_2$CN at each strain level. Bond lengths scale uniformly with the lattice parameters since fractional coordinates are fixed throughout.}
\label{stab:bonds}
\end{table}

\subsection{Bond-Projected Force Constants}
\label{sssec:fc}

To quantify the bond softening implied by the lattice elongation in Sec.~\ref{sssec:coords}, we projected the real-space interatomic force-constant tensor $\Phi_{\alpha\beta}(i,j)$ onto each bond's unit vector, $k_{\mathrm{bond}} = \hat{u}^{\mathsf{T}} \Phi \hat{u}$, using the force constants from \texttt{q2r.x}. Table~\ref{stab:fc} lists the resulting nearest-neighbor B--C and B--N force constants at each strain level. Both bonds soften monotonically with strain: the B--C force constant decreases by 47.7\% and the B--N force constant by 70.2\% across the full 0--6\% range, with B--N softening consistently faster than B--C at every strain step. The preferential softening of B--N drives the phonon softening and enhanced electron-phonon coupling reported in the main text.

\begin{table} [!ht]
\setlength{\tabcolsep}{10pt}
\centering
\begin{tabular}{lcccc}
\toprule
Bond & $\varepsilon=0\%$ & $\varepsilon=2\%$ & $\varepsilon=4\%$ & $\varepsilon=6\%$ \\
\midrule
B--C & 13.16 & 10.74 & 8.67 & 6.88 \\
B--N & 10.18 & 7.37 & 5.02 & 3.04 \\
\bottomrule
\end{tabular}
\caption{Nearest-neighbor bond-projected force constants (eV/\AA$^2$) in R$\bar{3}$m B$_2$CN at each strain level, obtained by projecting the real-space force-constant tensor onto the bond direction. Values are reported as magnitudes; the underlying tensor elements carry the restoring-force sign convention of the interatomic force-constant matrix.}
\label{stab:fc}
\end{table}

\section{Elastic Properties}
\label{ssec:elastic_supp}

\subsection{Born--Huang Stability Criteria}
\label{sssec:born_huang}

The Born--Huang stability conditions for trigonal crystals are given in Sec.~III of the main text. Table~\ref{stab:born_huang} lists their numerical values at each strain level, computed from the elastic constants in Table~I. All four conditions are satisfied with a large margin across the 0--6\% strain range, and no strain level approaches a mechanical instability.

\begin{table} [!ht]
\setlength{\tabcolsep}{10pt}
\centering
\begin{tabular}{lcccc}
\toprule
Condition & $\varepsilon=0\%$ & $\varepsilon=2\%$ & $\varepsilon=4\%$ & $\varepsilon=6\%$ \\
\midrule
$C_{11} - |C_{12}|$ (GPa)                               & 342.58 & 342.71 & 342.26 & 340.34 \\
$C_{44}$ (GPa)                                          & 292.97 & 293.06 & 292.72 & 291.20 \\
$(C_{11}+C_{12})C_{33} - 2C_{13}^{2}$ ($10^5$~GPa$^2$)  &   6.51 &   6.51 &   6.49 &   6.38 \\
$(C_{11}-C_{12})C_{44}$ ($10^5$~GPa$^2$)                &   1.00 &   1.00 &   1.00 &   0.99 \\
\bottomrule
\end{tabular}
\caption{Numerical evaluation of the Born--Huang stability conditions for trigonal crystals at each strain level. Conditions 1 and 2 are in GPa; conditions 3 and 4 are in $10^{5}$~GPa$^2$. All values are positive, confirming mechanical stability throughout the strain series.}
\label{stab:born_huang}
\end{table}

\subsection{Directional Elastic Properties from ELATE}
\label{sssec:elate}

\begin{figure} [!ht]
\centering
\includegraphics[width=\textwidth]{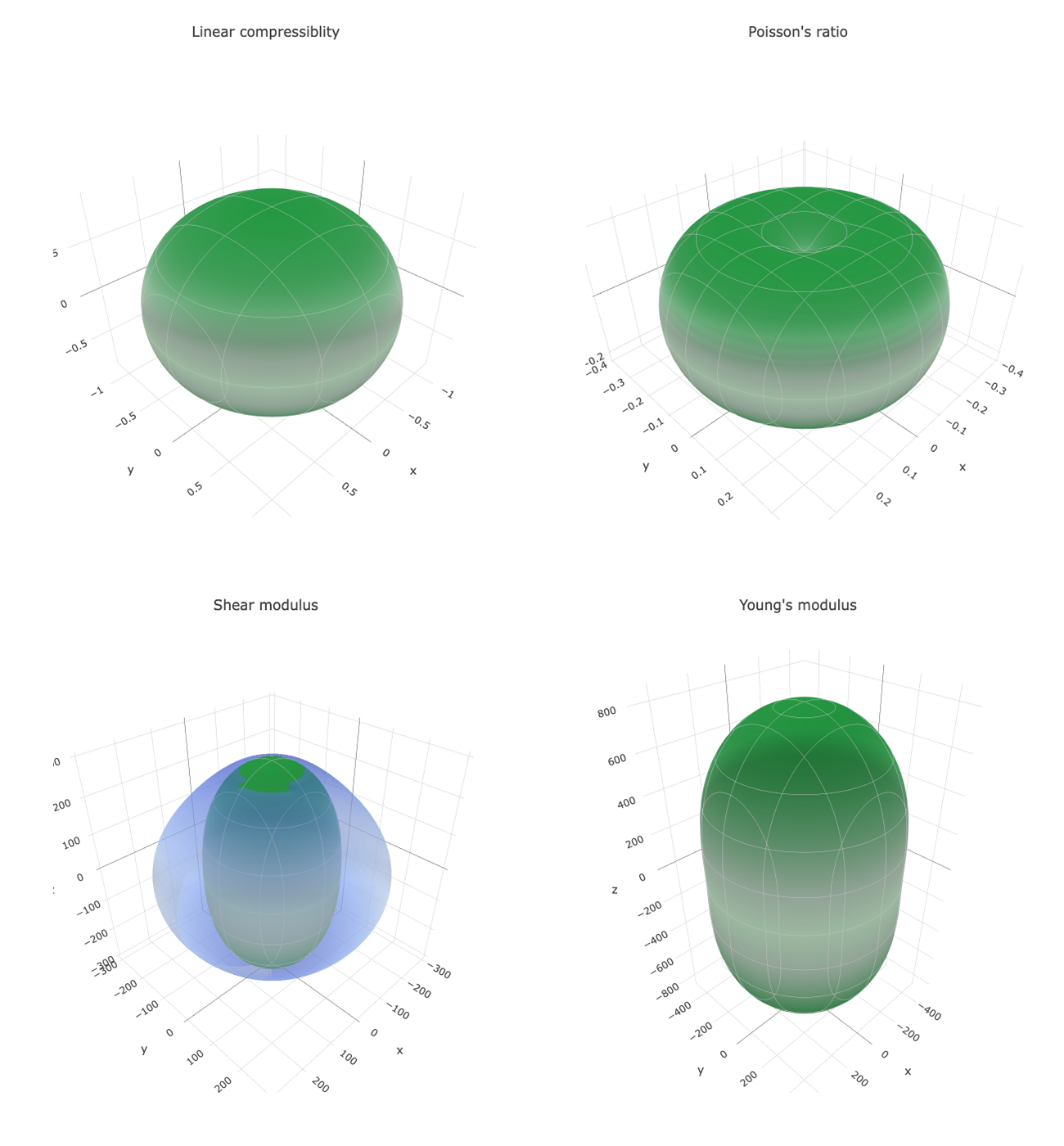}
\caption{Directional elastic properties of R$\bar{3}$m B$_2$CN at $\varepsilon = 0\%$ from ELATE~\cite{gaillac2016elate}. (a)~Young's modulus $E(\hat{n})$, (b)~linear compressibility $\beta(\hat{n})$, (c)~shear modulus $G(\hat{n})$, and (d)~Poisson's ratio $\nu(\hat{n})$.}
\label{sfig:elate_0}
\end{figure}

\begin{figure} [!ht]
\centering
\includegraphics[width=\textwidth]{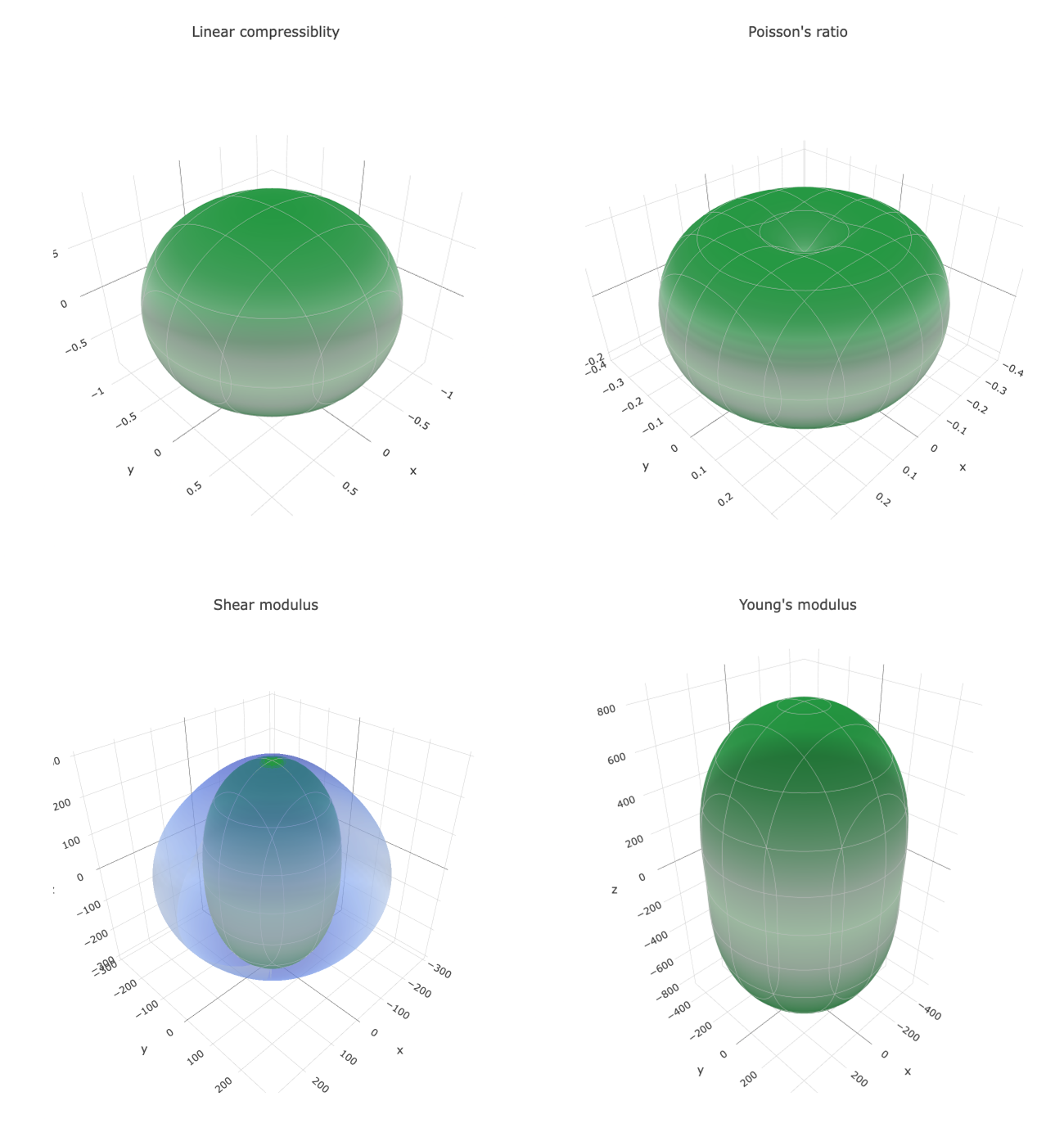}
\caption{Same as Fig.~\ref{sfig:elate_0} for $\varepsilon = 6\%$ isotropic tensile strain. The overall topology of each surface is preserved relative to the unstrained case, consistent with the negligible shift in $C_{ij}$ values across the strain series.}
\label{sfig:elate_6}
\end{figure}

The full angular dependence of the elastic response was computed with ELATE~\cite{gaillac2016elate}. Figures~\ref{sfig:elate_0} and~\ref{sfig:elate_6} show the directional Young's modulus $E(\hat{n})$, linear compressibility $\beta(\hat{n})$, shear modulus $G(\hat{n})$, and Poisson's ratio $\nu(\hat{n})$ at 0\% and 6\% strain. $E(\hat{n}) $ remains nearly spherical at both strain levels, consistent with the three-fold rotational symmetry of the rhombohedral lattice and the absence of any pronounced soft direction. The slight flattening along the $c$-axis in $G(\hat{n})$ reflects the smaller $C_{44}$ relative to $C_{11}$ reported in Table~I.

\section{Phonon Dispersions at 8\% and 10\% Strain}
\label{ssec:unstable_phonon}

\begin{figure}[]
\centering
\includegraphics[width=\columnwidth]{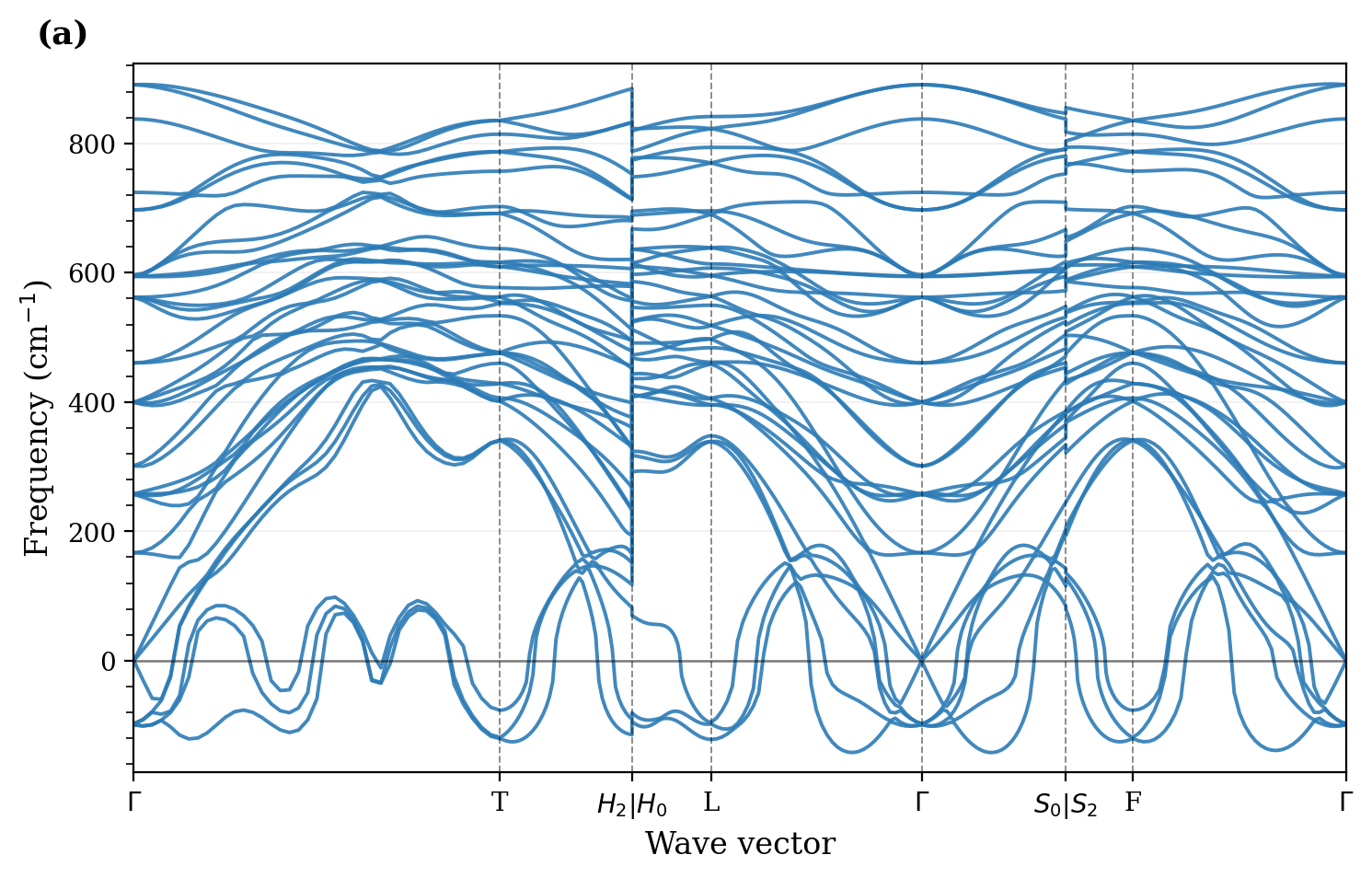}
\includegraphics[width=\columnwidth]{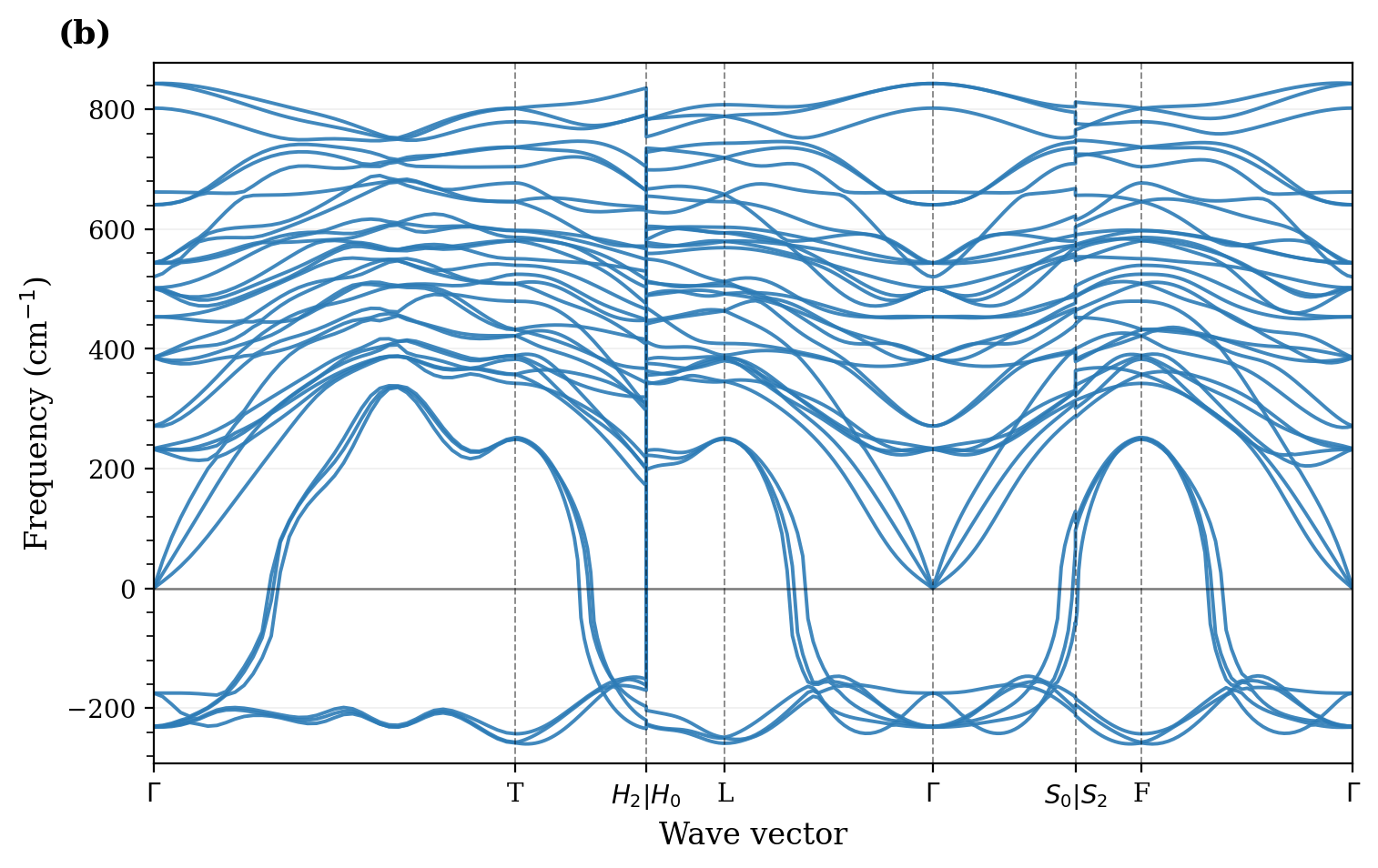}
\caption{Phonon dispersions of R$\bar{3}$m B$_2$CN at 8\% and 10\% isotropic tensile strain. Imaginary frequencies (plotted as negative values) confirm that 6\% is near the upper limit of dynamical stability.}
\label{sfig:phonon_unstable}
\end{figure}

Phonon calculations at 8\% and 10\% isotropic tensile strain reveal imaginary branches over broad portions of the Brillouin zone, placing the dynamical stability boundary just above 6\% strain (Sec.~VI of the main text). Figure~\ref{sfig:phonon_unstable} shows the full dispersion curves at both strain levels. Imaginary modes, plotted as negative frequencies, appear across the entire zone, consistent with a soft-mode instability from excessive weakening of the interatomic force constants.

\section{Fermi Surface Sheet Decomposition}
\label{ssec:fs_sheets}

\begin{figure}[]
\centering
\includegraphics[width=\columnwidth]{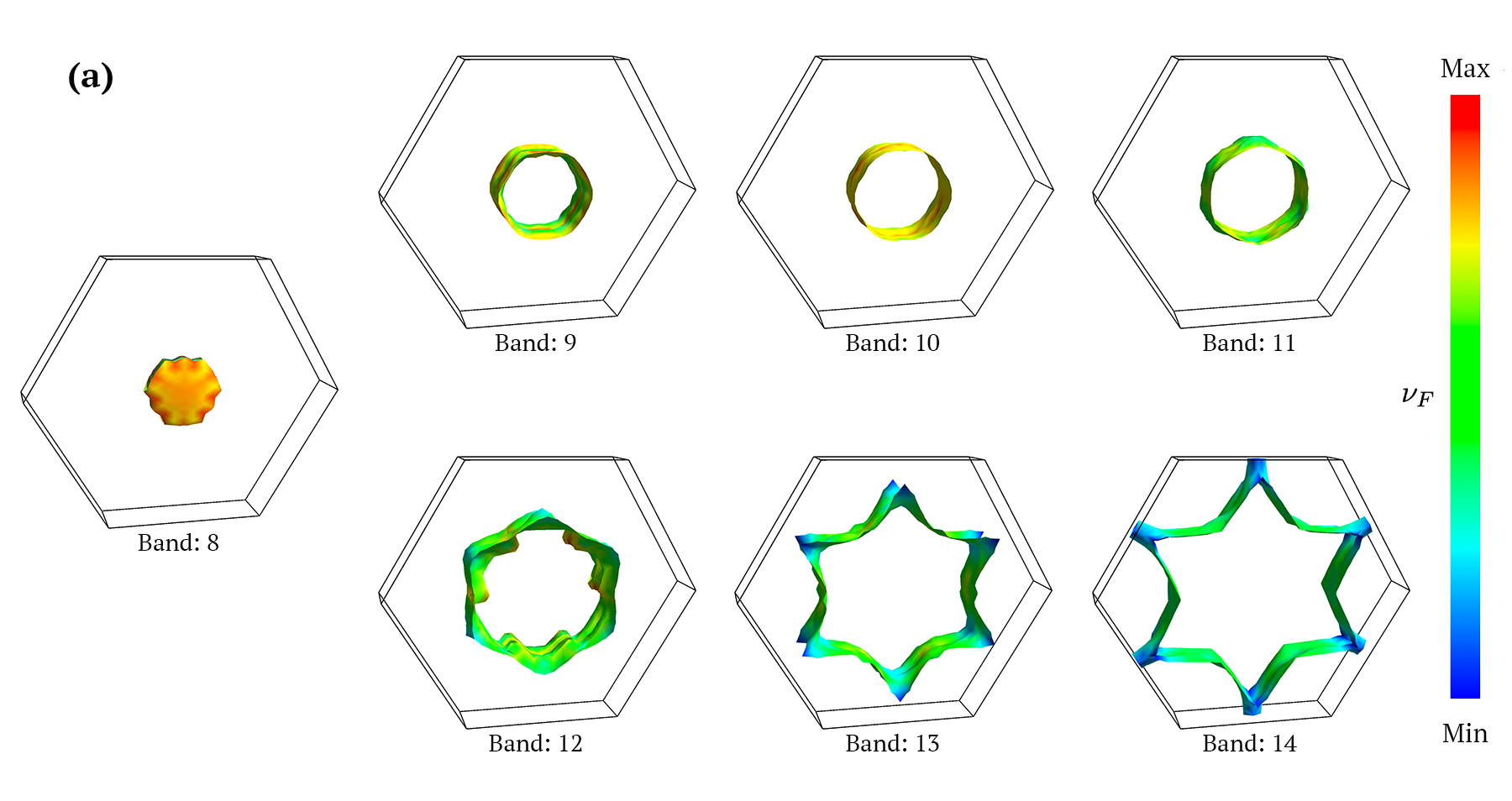}
\includegraphics[width=\columnwidth]{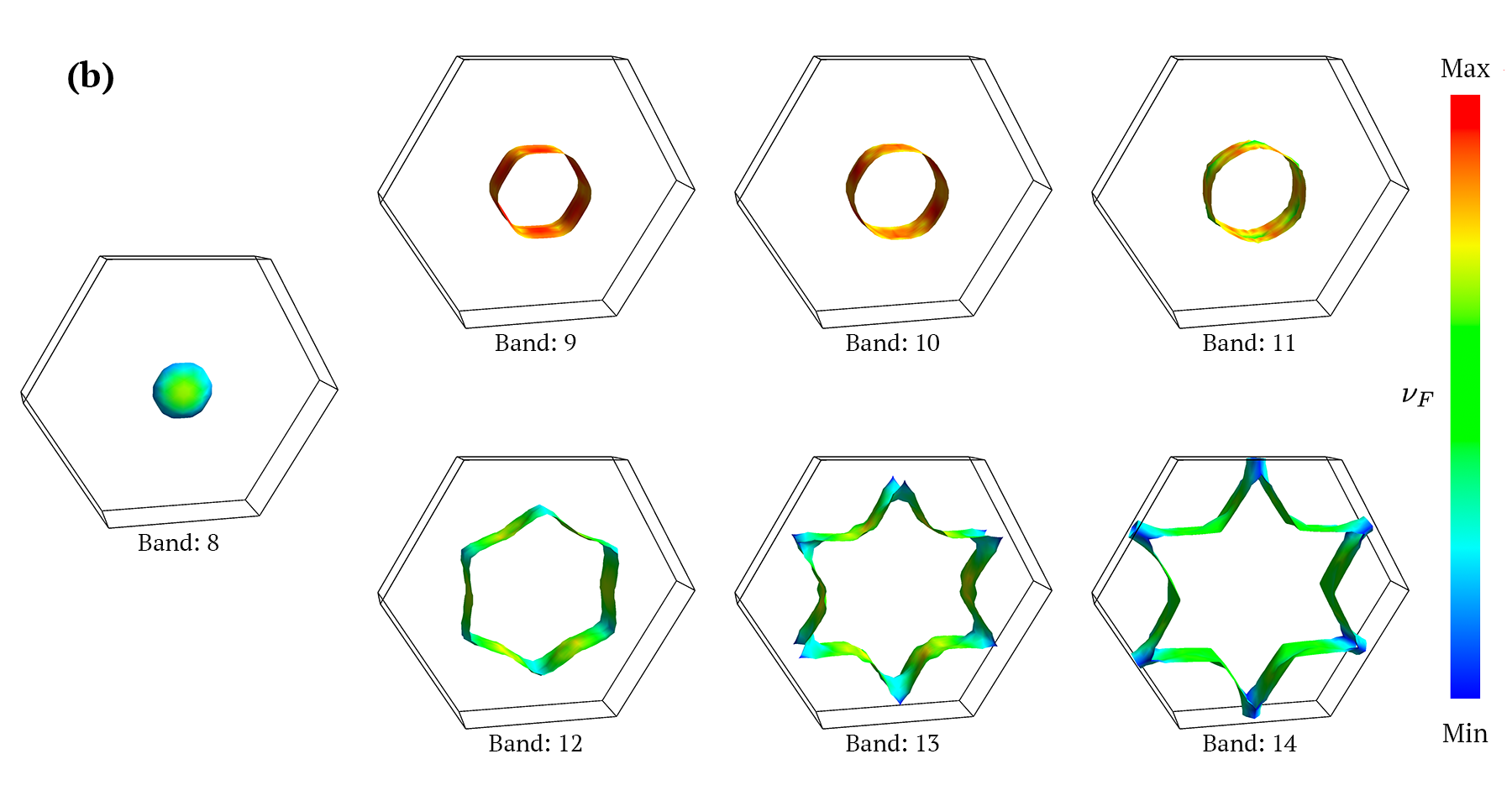}
\caption{Individual Fermi surface sheets from bands 8--14 for
R$\bar{3}$m B$_2$CN at (a)~0\% and (b)~6\% strain, colored by the magnitude of the Fermi velocity $v_F$.}
\label{sfig:fs_sheets}
\end{figure}

R$\bar{3}$m B$_2$CN has seven Fermi sheets (bands 8--14) at all strain levels. Figure~\ref{sfig:fs_sheets} shows each sheet at 0\% and 6\% strain, colored by the local Fermi velocity $v_F$. The total $N(E_F)$ at each strain level is given in Section~IV of the main text.

\section{Eliashberg Spectral Function: Supplementary Analysis}
\label{ssec:a2f_supp}

\subsection{Sensitivity to the Coulomb Pseudopotential $\mu^*$}
\label{sssec:mustar}

The Coulomb pseudopotential $\mu^* = 0.10$ used throughout is the standard value for covalent light-element systems. Table~\ref{stab:mustar} reports $T_c^{\rm AD}$ and $T_c^{\rm iso}$ for $\mu^* \in \{0.08, 0.10, 0.12, 0.15\}$ at 0\% and 6\% strain, using the \emph{isotropic} Eliashberg solver as a computationally cheaper proxy for testing $\mu^*$ sensitivity. This isotropic solver systematically underestimates $T_c$ relative to the fully anisotropic calculation reported in the main text (e.g., $T_c^{\rm iso} = 48.5$~K vs. $T_c^{\rm aniso} = 53.5$~K at 0\% strain and $\mu^*=0.10$), consistent with the anisotropy enhancement discussed in Sec.~VI.C. The $\mu^*$-dependence trend, rather than the absolute values, is the relevant result here.

\begin{table} [!ht]
\setlength{\tabcolsep}{10pt}
\centering
\begin{tabular}{lcccc}
\toprule
$\mu^*$ & $T_c^{\rm AD}$ (0\%) & $T_c^{\rm iso}$ (0\%) &
          $T_c^{\rm AD}$ (6\%) & $T_c^{\rm iso}$ (6\%) \\
\midrule
0.08        & 46.5 & 52.5 & 62.3 & 85.5 \\
0.10 (used) & 42.2 & 48.5 & 60.2 & 82.5 \\
0.12        & 38.0 & 45.5 & 58.0 & 80.5 \\
0.15        & 31.9 & 41.5 & 54.8 & 77.5 \\
\bottomrule
\end{tabular}
\caption{Dependence of $T_c^{\rm AD}$ and $T_c^{\rm iso}$ (K) on the Coulomb pseudopotential $\mu^*$ at $\varepsilon = 0\%$ and $6\%$. The adopted value $\mu^* = 0.10$ is marked. The liquid-nitrogen boiling point (77~K) is exceeded by $T_c^{\rm iso}$ at 6\% strain for all values of $\mu^*$ in the table.}
\label{stab:mustar}
\end{table}

\section{Anisotropic Eliashberg Equations: Additional Results}
\label{ssec:aniso_supp}

\subsection{Sheet-Resolved Gap Mapping}
\label{sssec:gap_sheet}

The gap distribution $\Delta_{nk} $ stratifies into distinct bands at high strain, as discussed in Sec.~VI.D of the main text, with each band mapping onto one of the seven Fermi sheets (bands 8--14). Table~\ref{stab:gap_sheet} reports the mean gap on each sheet normalized by $k_BT_c $, extracted from the anisotropic Eliashberg solution at the lowest computationally accessible temperature at each strain level. Normalization by $k_BT_c $ permits comparison across the strain series despite differing minimum temperatures. The BCS weak-coupling limit gives $\Delta_0/k_BT_c = 1.764 $ from $2\Delta_0 = 3.528\,k_BT_c $. The systematic enhancement above this value at 6\% strain follows from the strong-coupling regime ($\lambda = 2.620 $) established in Sec.~V. Band indices at $\varepsilon = 4\%$ are re-indexed from 1 in the EPW output and are mapped to the DFT band numbering (8-14) used throughout.

\begin{table} [!ht]
\setlength{\tabcolsep}{10pt}
\centering
\begin{tabular}{lcccc}
\toprule
Band    & $\varepsilon=0\%$ & $\varepsilon=2\%$ &
          $\varepsilon=4\%$ & $\varepsilon=6\%$ \\
(sheet) & (25~K)            & (40~K)            &
          (55~K)            & (55~K)            \\
\midrule
8        & 1.69 & 1.79 & 1.55 & 2.14 \\
9        & 1.71 & 1.77 & 1.80 & 2.09 \\
10       & 1.79 & 1.79 & 1.78 & 2.12 \\
11       & 1.76 & 1.93 & 1.76 & 2.22 \\
12       & 1.86 & 2.03 & 1.86 & 2.32 \\
13       & 1.86 & 1.99 & 1.71 & 2.30 \\
14       & 1.86 & 1.94 & 1.74 & 2.28 \\
\midrule
BCS ref. & \multicolumn{4}{c}{1.764} \\
\bottomrule
\end{tabular}
\caption{Sheet-resolved mean superconducting gap normalized by $k_BT_c$ for each Fermi sheet of R$\bar{3}$m B$_2$CN. Values are extracted from the anisotropic Eliashberg solution at the lowest computationally accessible temperature at each strain level: 25~K (0\%), 40~K (2\%), and 55~K (4\% and 6\%). The BCS weak-coupling limit $\Delta_0/k_BT_c = 1.764$ serves as a reference.}
\label{stab:gap_sheet}
\end{table}

\FloatBarrier

\bibliographystyle{ieeetr}
\bibliography{references}